\documentclass[preprint, aps]{revtex4}
\usepackage[british]{babel}
\usepackage{latexsym}
\usepackage{graphicx}
\usepackage{amsmath}
\begin{document}
\title{
Solution of a Braneworld Big Crunch/Big Bang Cosmology}
\author{Paul L. McFadden}
\email{p.l.mcfadden@damtp.cam.ac.uk}
\author{Neil  Turok}
\email{n.g.turok@damtp.cam.ac.uk}
\affiliation{DAMTP, Center for Mathematical Sciences, Wilberforce Road, 
  Cambridge, CB3 0WA, UK.}
\author{Paul J. Steinhardt}
\email{steinh@princeton.edu}
\affiliation{Joseph Henry Laboratories, Princeton University,
  Princeton, New Jersey 08544, USA.}

\date{\today}

\newcommand{\nc}{\newcommand}
\nc{\rnc}{\renewcommand}
\nc{\eg}{\textit{e.g. }}
\nc{\etal}{\textit{et al.}}
\nc{\dx}{\mathrm{d} ^4 x}
\nc{\D}{\partial}
\rnc{\d}{\mathrm{d}}
\nc{\gdxdx}{ g_{\mu \nu}(x) \mathrm{d}x^\mu \mathrm{d}x^\nu}
\nc{\ndxdx}{ \eta _{\mu \nu}(x) \mathrm{d}x^\mu \mathrm{d}x^\nu}
\nc{\tgdxdx}{\tilde{g}_{\mu \nu}(x) \mathrm{d}x^\mu \mathrm{d}x^\nu}
\nc{\dxdx}{\mathrm{d}x^\mu \mathrm{d}x^\nu}
\nc{\g}{g_{\mu \nu}}
\nc{\gpm}{g^\pm _{\mu \nu}}
\nc{\gp}{g^+ _{\mu \nu}}
\nc{\gm}{g^- _{\mu \nu}}
\rnc{\[}{\begin{equation}}
\rnc{\]}{\end{equation}}
\nc{\bea}{\begin{eqnarray}}
\nc{\eea}{\end{eqnarray}}
\nc{\ie}{\textit{i.e. }}
\nc{\iec}{\textit{i.e., }}
\nc{\etc}{\textit{etc. }}
\rnc{\tt}{\rightarrow} 
\rnc{\inf}{\infty}
\rnc{\L}{L}
\nc{\w}{\omega}
\nc{\inv}{\mathrm{inv}}
\nc{\tN}{\tilde{N}}
\nc{\tB}{\tilde{B}}
\nc{\tA}{\tilde{A}}
\nc{\tk}{\tilde{k}}
\nc{\cosech}{\mathrm{cosech}}
\nc{\sech}{\mathrm{sech}}
\nc{\tl}{\bar{\lambda}}
\nc{\tnu}{\tilde{\nu}}
\nc{\Nbar}{\bar{N}}
\begin{abstract}
We solve for the cosmological perturbations in a 
five-dimensional background consisting of
two separating or colliding boundary branes, as
an expansion in the collision speed $V$ divided by
the speed of light $c$.  Our 
solution permits 
a detailed check of the validity of 
four-dimensional effective theory in the 
vicinity of the event corresponding to
the big crunch/big bang singularity. We show that the
four-dimensional description fails at the first
nontrivial order in $(V/c)^2$. 
At this order, there is nontrivial  
mixing of the two relevant four-dimensional perturbation modes  
(the growing and decaying modes) as the 
boundary branes move from the narrowly-separated limit
described by Kaluza-Klein theory to the 
well-separated limit where gravity is confined to
the positive-tension brane. 
We comment on the cosmological
significance of the result and compute other quantities
of interest in five-dimensional cosmological scenarios.

\end{abstract}

\maketitle

\section{Introduction}

One of the most striking implications of string theory and
M theory is that there are extra spatial dimensions whose
size and shape determine the particle spectrum and couplings of the
low energy world. 
If the extra dimensions are compact and of fixed size $R$, their
existence results in a tower of Kaluza-Klein massive modes 
whose mass scale is set by $R^{-1}$. Unfortunately, this 
prediction is hard to test if the only energy scales 
accessible to experiment are much lower than $R^{-1}$.
At low energies, the massive modes
decouple from the low energy effective theory and are,
for all practical purposes, invisible. Therefore, we have no means
of checking whether the four-dimensional effective
theory observed to fit particle physics 
experiments is actually the outcome
of a simpler higher-dimensional theory.

The one situation where the extra dimensions seem
bound to reveal themselves is in cosmology. At the big bang,
the four-dimensional effective theory (Einstein gravity 
or its stringy generalization) breaks down, indicating 
that it must be replaced by an improved description.
There are already suggestions of improved behavior
in higher-dimensional string and M theory.
If matter is localized on two
boundary branes,
the matter density remains finite
at a brane collision even though this moment is, from the  
perspective
of the four-dimensional effective theory,  
the big bang singularity \cite{Cyclicevo,Ekpyrotic,Seiberg}. Likewise, 
the equations of motion for fundamental strings are
actually regular at $t=0$ in string theory, in the relevant
background solutions \cite{perry, Gustavo}.

In this paper, we shall not study the singularity itself. 
Instead, we will study the behavior of higher-dimensional
gravity as the universe emerges from a brane collision.
Our particular concern is to determine the extent to which
the four-dimensional
effective theory accurately captures the higher-dimensional
dynamics near the big bang singularity. 

The model we study is the simplest possible model 
of braneworld 
gravity. It consists
of two empty $Z_2$-branes (or orbifold planes)
of opposite tension separated by a five-dimensional
bulk with negative cosmological constant, 
corresponding to an anti-de Sitter (AdS) radius $L$ \cite{RSI}. 
Many works have been devoted to obtaining
exact or approximate solutions for this model,
for static or moving branes \cite{K&S, Toby, Gonzalo, SC1, SC2}. 
Our methods have much in common with these
earlier works, in particular the idea that, when the 
branes move slowly, the four-dimensional effective
theory works well. However, our focus and goals are rather
different. 

When the two boundary
branes are very close to one another, the warping of the
five-dimensional bulk and the tension of the branes
become irrelevant. In this situation, 
the low energy modes of the system are 
well-described by a simple Kaluza-Klein reduction
from five to four dimensions, \iec gravity plus a
scalar field (the $Z_2$ projections eliminate the
gauge field zero mode).  We shall verify this 
expectation.  However, when the two branes are widely 
separated, the physics is quite different.
In this regime, the warping of the
bulk plays a key role, causing  
the low energy gravitational modes to be localized 
on the positive-tension brane \cite{RSII,garriga,giddings}.
The four-dimensional effective theory describing this
new situation is nevertheless
identical, consisting of Einstein gravity and 
a scalar field, the radion, describing 
the separation of the two branes. 

In this paper, we study the transition between these 
two regimes -- from the naive Kaluza-Klein reduction to localized
Randall-Sundrum gravity -- at finite brane speed. 
In the two asymptotic regimes -- the narrowly-separated brane limit and
the widely-separated limit -- the cosmological 
perturbation modes show precisely the behavior predicted
by the four-dimensional effective theory. There are
two massless scalar perturbation modes; in 
longitudinal gauge, and in the long 
wavelength ($k\rightarrow 0$) limit, one mode 
is constant and the other
decays as $t_4^{-2}$, where $t_4$ is the  
conformal time.
In the four-dimensional description, 
these two perturbation modes are entirely distinct: 
one is the curvature perturbation mode; the
other is a local time delay to the big bang.
However, we shall show that in the five-dimensional
theory, at first nontrivial order in the speed of
the brane collision, the two modes mix.  If,
for example, one starts out in 
the time delay mode at small $t_4$, one ends up 
in a mixture of the time
delay and curvature perturbation modes as 
$t_4 \rightarrow \infty$. 
Thus the two cosmological perturbation modes --
the growing and decaying adiabatic modes -- mix in 
the higher-dimensional braneworld 
setup, a phenomenon which is prohibited
in four dimensions. 

The mode-mixing occurs as a result of the 
qualitative change in the nature of the 
low energy modes of the system. At small
brane separations the low energy modes are nearly
uniform across the extra dimension. Yet as the brane
separation becomes larger than the bulk warping scale,
the low energy modes
become exponentially localized on
the positive-tension brane. If the branes separate at
finite speed, the localization process fails to 
keep pace with the brane separation and the 
low energy modes do not evolve adiabatically.
Instead, they evolve into a mixture
involving higher Kaluza-Klein modes,
and the four-dimensional
effective description fails. 

As mentioned, the mixing we see
between the two scalar perturbation modes would
be prohibited in any local four-dimensional
effective theory consisting of Einstein gravity and matter
fields, no matter what the matter fields were. 
Therefore the mixing is a truly five-dimensional phenomenon, 
which cannot be modeled with a local
four-dimensional effective theory.
There is an independent argument
against the existence of any local four-dimensional 
description of these phenomena.  In standard 
Kaluza-Klein theory, it is well known that 
the entire spectrum of massive modes is 
actually spin two \cite{duff}. Yet, despite many attempts, no
satisfactory Lagrangian description of massive, purely spin
two fields has ever been found \cite{deser,damour}.
Again, this suggests that one should not expect to 
describe the excitation
of the higher Kaluza-Klein modes
in terms of an improved, local,
four-dimensional effective theory. 

The system we study consists of two branes emerging from
a collision. In this situation, 
there are important simplifications which allow
us to specify initial data rather precisely.
When the brane separation is small, the fluctuation modes 
neatly separate into light Kaluza-Klein zero modes, which are constant along
the extra dimension, and massive modes with nontrivial 
extra-dimensional dependence. Furthermore, the brane tensions
and the bulk cosmological constant become irrelevant 
at short distances. It is thus natural
to specify initial data which map precisely onto 
four-dimensional fields in the naive dimensionally-reduced 
theory describing the limit of  narrowly-separated branes.
 With initial data specified this way, there
are no ambiguities in the system. The two branes
provide boundary conditions for all time and the five-dimensional
Einstein equations yield a unique solution, for arbitrary
four-dimensional initial data.

Our main motivation is the study of cosmologies
in which the big bang was a brane collision, such as the 
cyclic model \cite{Cyclicevo}. Here, a period of 
dark energy domination, followed by slow contraction of the
fifth dimension, renders the branes locally flat and parallel 
at the collision. During the slow contraction phase, growing,
adiabatic, 
scale-invariant
perturbations are imprinted on the branes prior to the 
collision. However, if the system is accurately described
by four-dimensional effective theory throughout, 
then, as a number of authors have noted \cite{brand1, brand2, lyth1, jch1, jch2, Creminelli}, there is an apparent roadblock to the passage of 
the scale-invariant perturbations across the bounce.
Namely, it is hard to see how the growing mode
in the contracting phase, usually described as a local time delay, 
could match onto the growing mode in
the expanding phase, usually
described as a curvature perturbation. 
In this paper, we show that the 
four-dimensional effective theory fails at order $(V/c)^2$,
where $V$ is the collision speed and $c$ is the speed
of light. The four-dimensional description
works well when the branes are close together, or far
apart. However, as the branes move from one regime to the
other, the two four-dimensional modes mix
in a nontrivial manner.

The mixing we find is an explicit demonstration of a new
physical principle: namely, the approach of two boundary branes
along a fifth dimension produces physical effects that cannot 
properly be modeled by a local four-dimensional effective theory. In this paper, we deal
with the simplest case involving two empty
boundary branes separated by a bulk with a negative
cosmological constant. For the cyclic model, the details are
more complicated \cite{cyclicpaper}. The bulk possesses
an additional
bulk stress $\Delta T_5^5$, associated with the inter-brane force,
that plays a vital role in
converting a  growing mode 
corresponding to a pure time delay perturbation
into a mixture of time delay and curvature modes on the brane.  We also 
explain how subsequent five-dimensional effects cause 
cause the curvature on the brane to feed 
into the adiabatic
growing mode perturbation in the four-dimensional effective
theory after the bang.
The details are different, but the
general principle is the same as in the case considered
in this paper. One must go beyond
the four-dimensional effective theory 
and consider the full five-dimensional theory to 
compute properly  the evolution of perturbations before and 
after a brane collision.

Even though our main concern is with cyclic/ekpyrotic models,
our methods are likely to be more 
broadly applicable in braneworld
models. (See \eg \cite{langlois,maartens,durrer} 
for reviews). Our methods may be extended, for example, to models 
with better motivation from fundamental theory, such as
heterotic M theory \cite{Lukas&Ovrut,Lukas}.  
One may also include matter on the branes. 
Another interesting application would be 
to study the evolution of 
a four-dimensional black 
hole in the limit of narrowly-separated
branes, {\it i.e.}, a black string,
as the two branes separate and the Gregory-Laflamme 
instability appears. 

The outline of this paper is as follows.
In Section II we provide an overview of 
our three solution methods. In Section III
we solve for the background and cosmological
perturbations using a series expansion in time about the
collision.  In Section IV we present an improved method in
which the dependence on the fifth dimension is approximated
using a set of higher-order Dirichlet or Neumann polynomials. 
In Section V we develop an expansion about the small-$(V/c)$ scaling
solution, before
comparing our results with those of the four-dimensional effective
theory in Section VI.  We conclude with a discussion of mode-mixing in
Section VII.  
Detailed explicit solutions may be found in
the Appendices, and the Mathematica code implementing our calculations 
is available online \cite{Website}.

\section{Three solution methods}

In this section, we review the three solution methods
employed, noting their comparative merits.
For the model considered here, with no
dynamical bulk fields, 
there is a Birkhoff-like theorem guaranteeing the existence
of coordinates in which the bulk is static. It is easy
to solve for the background in these coordinates. However, the
motion of the branes 
complicates the
Israel matching conditions rendering
the treatment of perturbations difficult. It is preferable
to choose a coordinate system in which the branes
are located at fixed spatial coordinate $y=\pm y_0$,
and the bulk evolves with time. 

We shall employ a coordinate system in which 
the five-dimensional line element for the background 
takes the form
\[
\label{metrica}
\d s^2 = n^2(t,y) (-\d t^2 +t^2 \d y^2) + b^2(t,y) \d \vec{x}^2,
\]
where $y$ parameterizes the fifth dimension and $x^i$, $i=1,2,3$,
the three noncompact dimensions. 
Cosmological isotropy  excludes $\d t \,\d x^i$ or
$\d y \,\d x^i$ terms, and homogeneity ensures 
$n$ and $b$ are independent of $\vec{x}$. 
The $t,y$ part of the background metric may then
be taken to be conformally flat. One can write the metric
for two-dimensional Minkowski spacetime in Milne form so
that the branes are located at  $y=\pm y_0$ and 
collide at $t=0$. By expressing the metric in 
locally Minkowski coordinates, $T=t \cosh{y}$ and
$Y=t \sinh{y}$, one sees that the collision 
speed is $(V/c)= 
\tanh{2 y_0}$ and the 
relative rapidity of the collision
is $2y_0$.
As long as the bulk metric is regular
at the brane collision and possesses cosmological symmetry, 
the line element may
always be put into the 
form (\ref{metrica}). Furthermore,
by suitably rescaling coordinates one can choose 
$b(0,y)=n(0,y)=1$.

In order to describe  perturbations about this background, one 
needs to specify an 
appropriate gauge choice.
Five-dimensional
longitudinal gauge is particularly convenient \cite{Carsten}:
firstly, it is completely gauge-fixed; 
secondly, the brane trajectories are unperturbed in this gauge \cite{TTS},
so that the Israel matching conditions are relatively simple; and finally,
in the absence of anisotropic stresses, the traceless part
of the Einstein $G^i_j$ (spatial) 
equation yields a constraint amongst the
perturbation variables, reducing them from four to three. 
In light of these advantages, we will work in
five-dimensional longitudinal gauge for the entirety of this paper.

Our three solution methods are as follows:

\begin{itemize}
\item {\bf Series expansion in \bf\textit{t}}
\end{itemize}

The simplest solution method for the background is to 
solve for the metric functions $n(t,y)$ and
$b(t,y)$ as 
a series in powers of $t$ about $t=0$. At each order,
the bulk Einstein equations yield
a set of ordinary differential equations in $y$,
with the boundary
conditions provided by the Israel matching conditions.
These are straightforwardly solved. 
A similar series approach, involving powers of $t$ and
powers of $t$ times $\ln{t}$ 
suffices for the perturbations.
 
The series approach is useful at small times $(t/L)\ll 1$
since it provides the precise 
solution for the background plus generic perturbations, 
close to the brane collision,
for all $y$ and for any collision rapidity $y_0$.  It allows
one to uniquely specify four-dimensional 
asymptotic data as $t$ tends to
zero.
However, 
the series thus obtained fails to converge at quite modest times. 
Following the system to long times
requires a more sophisticated method. Instead of taking
$(t/L)$ as our expansion parameter, we want to use the 
dimensionless rapidity of the brane collision $y_0$,
and solve at each order in $y_0$.

\begin{itemize}
\item {\bf Expansion in Dirichlet/Neumann polynomials in \bf\textit{y}}
\end{itemize}

In this approach we represent the spacetime metric
in terms of variables obeying either Dirichlet or Neumann 
boundary conditions
on the branes. We then express these 
variables as series of Dirichlet or Neumann  
polynomials in $y$ and $y_0$, 
bounded at each subsequent order by an increasing power
of the collision rapidity $y_0$.  (Recall that the
range of the $y$ coordinate is bounded by $|y|\le
y_0$).
The coefficients in these expansions are
undetermined functions of $t$. By 
solving the five-dimensional Einstein equations perturbatively
in $y_0$, we obtain a series of ordinary differential equations in
$t$, which can then be solved exactly.
In this Dirichlet/Neumann polynomial expansion, the Israel boundary conditions on the branes
are satisfied automatically at every order in $y_0$, while the initial
data at small $t$ are provided by the previous
series solution method.

The Dirichlet/Neumann polynomial expansion method yields 
simple, explicit solutions for the background and perturbations
as long as $(t/L)$ is smaller than $1/y_0$. Since $y_0 \ll 1$,
this considerably 
improves upon the naive series expansion in $t$. However,
for $(t/L)$ of order $1/y_0$, the expansion fails because
the growth in the coefficients overwhelms the extra powers of 
$y_0$ at successive orders. Since $(t/L) \sim 1/y_0$ corresponds
to brane separations of order the AdS radius, the Dirichlet/Neumann polynomial
expansion method fails to describe the late-time behavior of
the system, and a third method is needed.

\begin{itemize}
\item {\bf Expansion about the scaling solution}
\end{itemize}
The idea of our third method is to start by identifying a scaling
solution, whose form is independent of $y_0$ for all
$y_0\ll 1$. This scaling solution is well-behaved for
all times and therefore a perturbation expansion in $y_0$ 
about this solution is similarly well-behaved, even at very late 
times. To find the scaling solution, we first 
change variables from
$t$ and $y$ to an equivalent set of 
dimensionless variables. The characteristic velocity of the system is
the brane speed at the collision, $V=c\tanh 2 y_0 \sim 2 c y_0$,
for small $y_0$, where we have temporarily restored the speed of light $c$.    
Thus we have the dimensionless time 
parameter $x = y_0 ct/L \sim V t/L$, of order the time 
for the branes to separate by one AdS radius. We also
rescale the 
$y$-coordinate by defining $\w = y/y_0$, whose range is 
$-1\leq \w\leq 1$, independent of the characteristic velocity. 

As we shall show, when re-expressed in these variables, 
for small $y_0$, 
the bulk Einstein equations become perturbatively 
\textit{ultralocal}: at each order in $y_0$ one only has to 
solve 
an ordinary differential equation in $\w$, with a source
term determined by time derivatives of lower order terms.
The original partial
differential equations reduce to an infinite series of 
ordinary differential
equations in $\w$ which are then easily solved order by order 
in $y_0$. 

This method, an expansion in $y_0$ about the scaling 
solution, is
the most powerful and may be extended to 
arbitrarily long times $t$ and
for all brane separations. As is well known for this model,
a Birkhoff-like theorem holds for backgrounds with 
cosmological symmetry. 
The bulk in between the two branes is just a slice of five-dimensional
Schwarzschild-AdS spacetime \cite{langlois,maartens,durrer}
within which the two branes move, with a virtual black hole
lying outside of the physical region, beyond the negative-tension 
brane.  
As time proceeds, the negative-tension brane becomes closer
and closer to 
the horizon of the Schwarzschild-AdS black hole. Even though 
its location in the Birkhoff-frame (static) coordinates
freezes (see Figure 1), its proper speed grows and the 
$y_0$ expansion fails. 
Nonetheless, by analytic continuation of our solution in $\w$ and $x$, 
we are able to circumvent this temporary breakdown of
the $y_0$ expansion and follow
the positive-tension brane, and the perturbations localized
near it, as they run off to the boundary of anti-de Sitter 
spacetime. 

\begin{figure}
\includegraphics[width=10cm]{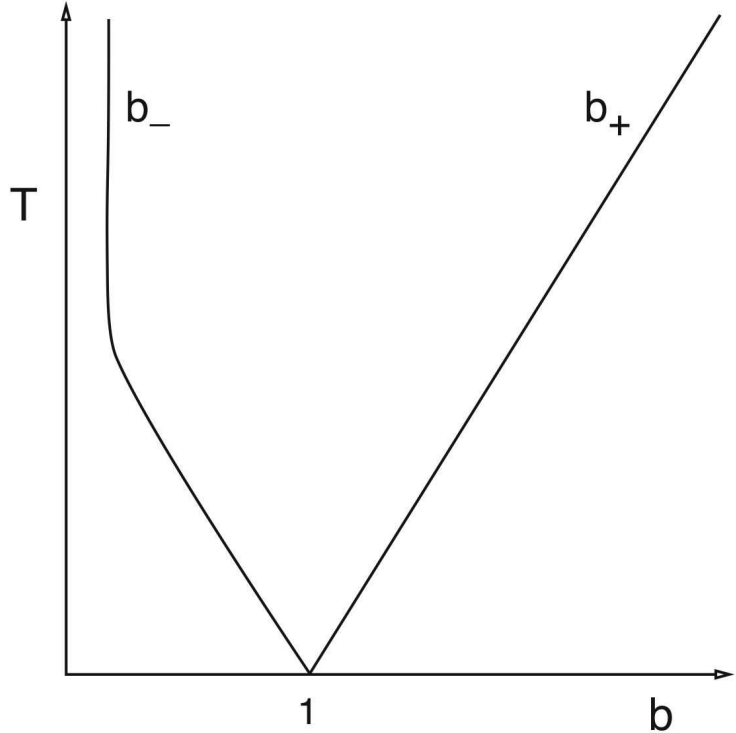} 
\caption{
The background brane scale factors $b_\pm$ plotted as a function of the Birkhoff-frame time $T$, where $b_{\pm}$ have been normalized to unity at $T=0$.  
In these coordinates the bulk is Schwarzschild-AdS: the
brane trajectories are then determined by integrating the Israel
matching conditions, given in Appendix E.  In the limit as $T\tt \inf$, the
negative-tension brane asymptotes to the event horizon of the black
hole, while the positive-tension brane asymptotes to the
boundary of AdS.
}


\end{figure}

Our expansion about the scaling solution is closely related to 
derivative-expansion techniques developed earlier by a number of authors
\cite{Toby, K&S, Gonzalo}.  In these works, an expansion in terms of
brane curvature over bulk curvature was used. For cosmological
solutions, this  
is equivalent to an expansion in $L \mathcal{H}^+$, where
$\mathcal{H}^+$ is the Hubble constant on the positive-tension brane.  
However, we specifically want to
study the time-dependence of the perturbations for all
times from the narrowly-separated to the well-separated brane limit.
For this purpose 
it is better to use a time-independent expansion
parameter ($y_0$) and  to
include all the appropriate time-dependence order by
order in the expansion. 

Moreover, in the earlier works, the goal was to find the
four-dimensional effective description more generally, 
without specifying that the branes emerged from a collision
with perturbations in the lowest Kaluza-Klein modes. 
Consequently, the solutions obtained
contained a number of undetermined functions.  
In the present context, however, the initial conditions along the
extra dimension are completely specified close to the brane collision
by the requirement that only the lowest Kaluza-Klein modes be excited.
The solutions we obtain here are fully determined, with no
arbitrary functions entering our results. 

Returning to the theme of the four-dimensional effective
theory, 
we expect on general grounds that this should be valid
in two particular limits:
firstly, as we have already discussed, a Kaluza-Klein
description will apply at early times near to the collision,
when the separation of the branes is much less than $L$. 
Here, the warping of the bulk geometry and the brane tensions can
be neglected.  Secondly, 
when the branes are separated by many AdS lengths, 
one expects gravity to become localized on the positive-tension
brane, which moves ever more slowly as time proceeds, so
the four-dimensional
effective theory should become more and more accurate.

Equipped with our five-dimensional solution for the background and
perturbations obtained by expanding about the scaling solution, we find ourselves
able to test the four-dimensional effective theory explicitly.  
We will show that the four-dimensional 
effective theory accurately
captures the five-dimensional dynamics to leading order in the
$y_0$-expansion, but fails at the first nontrivial order.
Our calculations reveal that
the four-dimensional perturbation modes undergo a mixing in the
transition between the Kaluza-Klein effective theory at early times
and the brane-localized gravity at late times.
This effect is a consequence of the momentary breakdown of the effective
theory when the brane separation is of the order of an AdS length, and
cannot be seen from four-dimensional effective theory calculations
alone. 

\section{Series solution in \lowercase{{\it \textbf{t}}}}

As described above,
we find it simplest to work in coordinates
in which the brane locations are fixed but the bulk
evolves.  The bulk metric is therefore given by (\ref{metrica}), with
the brane locations fixed at $y=\pm y_0$ for all time $t$.
The five-dimensional solution then has to satisfy both
the Einstein equations and the Israel matching conditions
on the branes \cite{Israel}. 

The bulk Einstein equations read $G_\mu^\nu = -\Lambda \delta_\mu^\nu$,
where the bulk cosmological constant
is $\Lambda = -6/L^2$ (we work in units in which the four-dimensional
gravitational coupling $8\pi G_4 = 8\pi G_5/L =1$). 
Evaluating the linear combinations $G^0_0 + G^5_5$
and $G^0_0 + G^5_5 - (3/2)G^i_i$ (where $0$ denotes time, $5$ labels 
the $y$ direction, and $i$ denotes one of the noncompact directions, 
with no sum implied), we find: 
\bea
\label{bgdeqn1}
\beta_{,\tau\tau}-\beta_{,y y} +\beta_{,\tau}^2-\beta_{,y}^2 +
12\,e^{2\nu} &=&0 \\
\label{bgdeqn2}
\nu_{,\tau\tau}- \nu_{,y y} 
+ \frac{1}{3}(\beta_{,y}^2-\beta_{,\tau}^2) - 2\,e^{2\nu} &=& 0,
\eea
where $(t/L)= e^\tau$, $\beta\equiv 3\ln{b}$ and $\nu \equiv  \ln{(nt/L)}$. 
The Israel matching conditions on the branes read \cite{Carsten,
TTS}
\[
\frac{b_{,y}}{b} = \frac{n_{,y}}{n} = \frac{nt}{\L},
\label{ibd}
\]
where all quantities are to be evaluated at the brane locations,
$y=\pm y_0$. 

We will begin our assault on the bulk geometry by constructing a
series expansion in $t$ about the collision, implementing the Israel
matching conditions on the branes at each order in $t$.
This series expansion in $t$ is then exact in both $y$ and the collision
rapidity $y_0$. 
It chief purpose will be to provide
initial data for the more powerful solution
methods that we will develop in the following sections.  

The Taylor series solution in $t$ 
for the background was first presented in \cite{TTS}:
\bea
n &=& 1 + (\sech\,{y_0}\sinh{y})\,\frac{t}{L}+\frac{1}{4}\,\sech^2\,{y_0}(-3+2\cosh{2y}
+\cosh{2y_0})\, \frac{t^2}{L^2} +O\Big(\frac{t^3}{L^3}\Big) \\
b &=& 1 +  (\sech\,{y_0}\sinh{y})\,\frac{t}{L}+\frac{1}{2}\,\sech^2\,{y_0}(\cosh{2y}
-\cosh{2y_0})\, \frac{t^2}{L^2} +O\Big(\frac{t^3}{L^3}\Big).
\eea
(Note that in the limit as $t\tt 0$ we correctly recover compactified Milne spacetime).

Here, however, we will need the perturbations as well.  
Working in five-dimensional longitudinal gauge for the reasons given
in the previous section, 
the perturbed bulk metric takes the form (see Appendix A)
\[
\d s^2 = n^2\left(-(1+2\Phi_L)\,\d t^2-2W_L\,\d t\d y+t^2\,(1-2\Gamma_L)\,\d
  y^2\right)+b^2\left(1-2\Psi_L\right)d\vec{x}^2,
\]
with $\Gamma_L=\Phi_L-\Psi_L$ being imposed by the five-dimensional
traceless $G^i_j$ equation.
The Israel matching conditions at $y=\pm y_0$ then read
\[
\Psi_{L\, ,y}=\Gamma_L \frac{nt}{L}, \qquad
\Phi_{L\, ,y}=-\Gamma_L \frac{nt}{L}, \qquad
W_L=0.
\label{lgi}
\]
Performing a series expansion, we find
\bea
\Phi_L &=& -\frac{B}{t^2}+\frac{B\,
  \sech{\,y_0}\sinh{y}}{t}+\Big(A-\frac{B}{8}-\frac{Bk^2}{4}+\frac{1}{6}B k^2\ln{|k t|}
\nonumber \\ && +\frac{1}{16}B\cosh{2y}\,(-1+6 \,y_0 \coth{2y_0})\,\sech^2{y_0}
-\frac{3}{8}B\,\sech^2{y_0}\sinh{2y}\Big) +O(t)\\
\Psi_L
&=&-\frac{B\,\sech{\,y_0}\sinh{y}}{t}+\Big(2A-\frac{B}{4}
+\frac{1}{3}Bk^2\ln{|kt|}+\frac{1}{4}B\cosh{2y}\,\sech^2{y_0}\Big)+O(t)\\
W_L &=& -\frac{3}{4}B\,\sech^2{y_0}\big(y\cosh{2y}-y_0\cosh{2y_0}\sinh{2y}\big)\,t+O(t^2),
\eea
where we have set $L=1$ (to restore $L$, simply replace $t\tt t/L$ and
$k\tt kL$).
Except for a few specific instances, we will now adopt this convention
throughout the rest of this paper.
The two arbitrary constants $A$ and $B$ (which may themselves be arbitrary 
functions of $\vec{k}$) have been chosen so that,
on the positive-tension brane, to leading order in $y_0$, $\Phi_L$ goes as
\[
\Phi_L = A - \frac{B}{t^2} + O(y_0)+O(k^2)+O(t).
\]

\section{Expansion in Dirichlet/Neumann polynomials}

\subsection{Background}

Having solved the relevant five-dimensional Einstein equations as
a series expansion in the time $t$ 
before or after the collision event, we now have
an accurate description of the behavior of the bulk at small $t$
for arbitrary collision rapidities.
However, in order to match onto the incoming
and outgoing states, 
we really want to study the
long-time behavior of the solutions, as
the branes become widely separated.  Ultimately this will
enable us to successfully map the system 
onto an appropriate four-dimensional effective description.
Instead of expanding in powers of the time, 
we approximate the five-dimensional solution
as a power series in the rapidity of the collision, and
determine each metric coefficient for all time
 at each order in the rapidity.

Our main idea is to express
the metric as a series of Dirichlet or Neumann polynomials in 
$y_0$ and $y$, bounded at order $n$ by a constant times
$y_0^n$, 
such that the series satisfies 
the Israel matching
conditions exactly at every order in $y_0$.
To implement this, we first change variables from
$b$ and $n$  to those obeying Neumann
boundary conditions. From 
(\ref{ibd}), $b/n$ 
is Neumann. 
Likewise, if we define $N(t,y)$ by 
\[
nt = \frac{1}{N(t,y) - y},
\label{nt}
\]
then one can easily check that $N(t,y)$ 
is also Neumann on the branes. Notice that if
$N$ and $b/n$ are constant, the metric (\ref{metrica})
is just that for anti-de Sitter spacetime. For fixed $y_0$, $N$ describes the
the proper separation of the
two branes, and $b$ is an additional
modulus describing the three-dimensional
scale-factor of the branes. 

Since $N$ and $b/n$ obey Neumann boundary conditions on
the branes, we
can expand both in a power series
\[
\label{Bgd_ansatz}
N= N_0(t)+\sum_{n=3}^\infty N_n(t) P_n(y), \qquad 
b/n=q_0(t)+\sum_{n=3}^\infty q_n(t) P_n(y), 
\]
where $P_n(y)$ are polynomials
\[
P_n(y)= y^n-\frac{n}{n-2} \,y^{n-2} \,y_0^2, \qquad n=3,4,\dots
\]
satisfying Neumann boundary conditions and
each bounded by $|P_n(y)|<2y_0^n/(n-2)$, for the relevant
range of $y$. Note that the time-dependent coefficients in this ansatz 
may also be expanded as a power series in $y_0$.
By construction, our ansatz
satisfies the Israel matching conditions exactly at each order
in the expansion.  The bulk
Einstein equations are not satisfied exactly, 
but as the expansion is
continued, the error terms are 
bounded by increasing powers of $y_0$.

Substituting the series ans{\"a}tze (\ref{Bgd_ansatz}) 
into the background Einstein equations (\ref{bgdeqn1}) and (\ref{bgdeqn2}),
we may determine the solution order by order in the rapidity $y_0$.
At each order in $y_0$, one generically obtains a number of linearly
independent algebraic equations, and at most one ordinary differential
equation in $t$.  The solution of the latter introduces a number of
arbitrary 
constants of integration into the solution.

To fix the arbitrary constants, one first applies the remaining
Einstein equations, allowing a small number to be eliminated.  
The rest are then determined using the series expansion in $t$
presented in the previous section: as this solution is
exact to all orders in 
$y_0$, we need only to expand it out to the relevant order in $y_0$,
before 
comparing it term by term with our Dirichlet/Neumann 
polynomial expansion (which is exact
in $t$ but perturbative in $y_0$), taken to a corresponding
order in $t$.
The arbitrary constants are then chosen so as to ensure the
equivalence of the two expansions in the region where both $t$ and
$y_0$ are small. 
This procedure suffices to fix all the remaining arbitrary constants.

The first few terms of the solution are
\bea
N_0 &=& {1\over t}-\frac{1}{2}\,t y_0^2+{1\over 24}\,t(8-9t^2)y_0^4+\dots \\
N_3 &=& -{1\over 6}+\left({5\over 72}-2t^2\right)y_0^2+\dots 
\eea
and
\bea
q_0 &=& 1 -\frac{3}{2}\,t^2 y_0^2+\left(t^2-\frac{7}{8}\,t^4\right)y_0^4+\dots \\
q_3 &=& -2\,t^3 y_0^2+\dots,
\eea
and the full solution up to $O(y_0^{10})$ may be found in Appendix B.


\subsection{Perturbations}

Following the same principles used in our treatment of the background,
we construct the two linear combinations 
\[
\label{phi4xi4}
\phi_4 = \frac{1}{2}(\Phi_L+\Psi_L), \qquad \xi_4= b^2
(\Psi_L-\Phi_L)=b^2 \Gamma_L,
\]
both of which obey Neumann boundary conditions on the branes, as may
be checked from (\ref{ibd}) and (\ref{lgi}).  In addition, $W_L$ already
obeys simple Dirichlet boundary conditions.

The two Neumann variables, $\phi_4$ and
$\xi_4$, are then expanded in a series
of Neumann polynomials and $W_L$ is 
expanded in a series of
Dirichlet polynomials, 
\[
D_n(y)=y^n-y_0^n,\qquad n =2,4,\dots, \qquad D_n(y)=y\,D_{n-1}(y),\qquad n=3,5,\dots,
\]
each bounded by $|D_n(y)|<y_0^n$ for $n$ even
and $y_0^n (n-1)/n^{n/(n-1)}$ for $n$ odd, over the relevant range
of $y$.  As in the case of the background, the time-dependent
coefficients multiplying each of the polynomials should themselves be
expanded in powers of $y_0$.

To solve for the perturbations it is sufficient to use only three of
the perturbed Einstein equations (any solution obtained may then be 
verified against the remainder).  
Setting
\bea
\Phi_L &=& \phi\, e^{-2\nu-\beta/3} \\
\Psi_L &=& \psi\, e^{-\beta/3} \\
W_L    &=& w\, e^{\tau-2\nu-\beta/3}
\eea
where $t=e^\tau$, $\beta=3\ln{b}$ and $\nu = \ln{nt}$,
the $G^5_i$, $G^0_i$ and $G^i_i$ equations take the form
\bea
\label{pe1}
w_{,\tau} &=& 2\,\phi_{,y} - 4\,e^{3\nu/2}\,(\psi\, e^{\nu/2})_{,y} \\
\label{pe2}
\phi_{,\tau} &=& {1\over 2}\,w_{,y} - e^{3\nu}\,(\psi\, e^{-\nu})_{,\tau} \\
\label{pe3}
(\psi_{,\tau}\, e^{\beta/3})_{,\tau} &=& (\psi_{,y}\, e^{\beta/3})_{,y}+
\psi\,e^{\beta/3}\,
\left(\frac{1}{3}\,\beta_{,\tau}^2-\frac{1}{9}\,\beta_{,y}^2-k^2\,e^{2(\nu-\beta/3)}\right)
\nonumber \\
&& -\frac{2}{9}\,e^{-2\nu+\beta/3}\,\left(\phi
\,(\beta_{,\tau}^2+\beta_{,y}^2)-
w\, \beta_{,\tau}\,\beta_{,y}\right) .
\eea

Using our Neumann and Dirichlet ans{\"a}tze for $\phi_4$, $\xi_4$ and
$W_L$, the Israel matching conditions are automatically satisfied and
it remains only to solve (\ref{pe1}), (\ref{pe2}) and (\ref{pe3})
order by order in the rapidity.
The time-dependent coefficients for $\phi_4$,
$\xi_4$ and  $W_L$ are then found to obey simple
ordinary differential equations, with solutions comprising
Bessel functions in $kt$, given in Appendix C.   
Note that it is not necessary for the set of Neumann or Dirichlet
polynomials we have used to be orthogonal to each other: 
linear independence is perfectly sufficient to determine all the
time-dependent coefficients order by order in $y_0$.

As in the case of the background, the arbitrary constants of
integration remaining in the solution after the application of the
remaining Einstein equations are fixed 
by performing a series
expansion of the solution in $t$.  This expansion can be compared
term 
by term with the series expansion in $t$ given previously, after
this latter series has itself been expanded in $y_0$.  
The arbitrary constants are then chosen so that the two expansions 
coincide in the region where both $t$ and $y_0$ are small.
The results of these calculations, at long wavelengths, are:
\bea
\Phi_L &=& A-B\left({1\over  t^2}-{k^2\over 6} {\rm ln}|kt|\right)
 +\left(A t +\frac{B}{t}\right)\,y + 
\dots \\
\Psi_L &=& 2 A +B {k^2\over 3} {\rm ln}|kt| -\left(A t+ \frac{B}{t}\right)\,y + \dots \\
W_L&=& 6 A\, t^2\, (y^2-y_0^2)+\dots 
\label{results}
\eea
where the constants A and B can be arbitrary functions of $k$. The solutions for
all $k$, to fifth order in $y_0$, are given in Appendix C.


\section{Expansion about the scaling solution}

It is illuminating to recast the results of the preceding sections in terms
of a set of dimensionless variables.  Using the relative velocity of
the branes at the moment of collision, $V= 2c \tanh y_0 \simeq 2c y_0$
(where we have temporarily re-introduced the speed of light $c$),
we may construct the dimensionless time parameter $x = y_0 ct/L \sim
Vt/L$ and the dimensionless $y$-coordinate $\w = y/ y_0 \sim y(c/V)$.

Starting from the full Dirichlet/Neumann 
polynomial expansion for the background given in Appendix
B, restoring $c$ to unity and setting $t=x L/y_0$ and $y = \w y_0$, 
we find that
\bea
\label{ad_n_series}
n^{-1} &=& \tN(x)-\w x + O(y_0^2) \\
\label{ad_q}
{b\over n} &=& q(x) + O(y_0^2),
\eea
where 
\bea
\label{N_series}
\tN(x) &=& 1 - \frac{x^2}{2} - \frac{3\,x^4}{8} - \frac{25\,x^6}{48} - \frac{343\,x^8}{384} - 
  \frac{2187\,x^{10}}{1280}+O(x^{12}) \\
\label{q_series}
q(x) &=& 1 - \frac{3\, x^2}{2} - \frac{7\, x^4}{8} - \frac{55\,
  x^6}{48} - \frac{245\, x^8}{128} - \frac{4617\, x^{10}}{1280} +O(x^{12}).
\eea
The single term in (\ref{ad_n_series}) linear in $\w$ is necessary in order that $n^{-1}$
satisfies the correct boundary conditions.  Apart from this 
one term, however, we see that to lowest order in $y_0$ the
metric functions above turn out to be completely independent of $\w$.  
Similar results are additionally found for the perturbations.

Later, we will see how this behavior leads to the emergence of
a four-dimensional effective theory.  For now, the key point to
notice is that this series expansion 
converges only for $x \ll 1$, corresponding to times $t \ll L/y_0$.
In order to study the behavior of the theory for all times therefore, we
require a means of effectively resumming the above perturbation expansion to all
orders in $x$.
Remarkably, we will be able to accomplish just this.  
The remainder of this section, divided into five parts, details our method and results:
first, we explain how to find and expand about the scaling solution,
considering only the background for simplicity.  
We then analyze various aspects of the background scaling solution,
namely, the brane geometry and the analytic continuation required to go
to late times, before moving on to discuss higher-order terms in the expansion.  
Finally, we extend our treatment to cover the perturbations.

\subsection{Scaling solution for the background}

The key to the our method is the observation that the 
approximation of small collision rapidity ($y_0\ll 1$) 
leads to a set of equations that are perturbatively ultralocal:
transforming to the dimensionless coordinates $x$ and $\w$, the
Einstein equations for the background
(\ref{bgdeqn1}) and (\ref{bgdeqn2}) become
\bea
\label{e1}
\beta_{,\w\w}+\beta_{,\w}^2-12\,e^{2\tnu}&=& y_0^2 \left(
x(x\beta_{,x})_{,x}+x^2\beta_{,x}^2\right) \\
\label{e2}
\tnu_{,\w\w}-\frac{1}{3}\,\beta_{,\w}^2+2\,e^{2\tnu} &=& y_0^2\big(
x(x\tnu_{,x})_{,x}-\frac{1}{3}\,x^2\beta_{,x}^2\big),
\eea
where we have introduced $\tnu = \nu+\ln{y_0}$.  
Strikingly, all the 
terms involving
$x$-derivatives are 
now suppressed by a factor of $y_0^2$ relative to the remaining terms.
This segregation of $x$- and $\w$-derivatives has profound
consequences: when solving perturbatively in $y_0$,  
the Einstein equations 
(\ref{e1}) and (\ref{e2}) reduce to 
a series of {\it ordinary} differential equations in $\w$,
as opposed to the partial differential equations we started off with.

To see this, consider expanding out both the Einstein equations (\ref{e1}) and (\ref{e2})
as well as the metric functions $\beta$ and $\tnu$ as a series in
positive powers of $y_0$.
At zeroth order in $y_0$, the right-hand sides of (\ref{e1}) and
(\ref{e2}) vanish, and the
left-hand sides can be integrated with respect to $\w$ to yield
anti-de Sitter space.  
(This was our reason for using $\tnu = \nu+\ln{y_0}$ rather than $\nu$:
the former serves to pull the necessary exponential term deriving from
the cosmological constant down to zeroth order in $y_0$, yielding 
anti-de Sitter space as a solution at leading order.  
As we are merely adding a
constant, the derivatives of $\tnu$ and $\nu$ are identical.)
The Israel matching conditions on the branes (\ref{ibd}), which in these coordinates read 
\[
\label{ibd2}
\frac{1}{3}\,\beta_{,\w}=\tnu_{,\w}=e^{\tnu},
\]
are not however sufficient to fix all the arbitrary functions of $x$ arising
in the integration with respect to $\w$.  In fact, two arbitrary functions of
$x$ remain in the solution, which may be regarded as time-dependent moduli describing the 
three-dimensional scale factor of the branes and their proper separation.
These moduli may be determined with the help of the $G^5_5$ Einstein equation as we will 
demonstrate shortly.  

Returning to (\ref{e1}) and (\ref{e2}) at $y_0^2$ order now,
the left-hand sides amount to ordinary differential equations in
$\w$ for the $y_0^2$ corrections to $\beta$ and $\tnu$.
The right-hand sides can no longer be neglected, but, because of the
overall factor of $y_0^2$, only the
time-derivatives of $\beta$ and $\tnu$ at {\it zeroth} order in $y_0$
are involved.
Since $\beta$ and $\tnu$ have already been determined to this order,
the right-hand sides therefore act merely as known source terms.
Solving these ordinary differential equations then introduces
two further arbitrary functions of $x$; these serve as $y_0^2$
corrections to the time-dependent moduli and may be fixed in the same
manner as previously.

Our integration scheme therefore proceeds at each order in $y_0$ via a
two-step process: first, we integrate the Einstein 
equations (\ref{e1}) and (\ref{e2}) to determine the $\w$-dependence of the bulk geometry, and then 
secondly, we fix the $x$-dependent moduli pertaining to the brane geometry using the $G^5_5$ equation.
This latter step works as follows: evaluating the $G^5_5$ equation on the branes, we can use the Israel matching
conditions (\ref{ibd2}) to replace the single $\w$-derivatives that appear in this equation, yielding
an ordinary differential equation in time for the geometry on each brane.
Explicitly, we find
\[
\left(\frac{bb_{,x}}{n}\right)_{,x}=0,
\]
where five-dimensional considerations (see Section VI) further allow us to fix the
constants of integration on the ($\pm$) brane as
\[
\label{G55}
\frac{b b_{,x}}{n}=\frac{b b_{,t}}{y_0
  n}=\frac{b_{,t_\pm}}{y_0} 
=\pm\frac{1}{y_0}\tanh{y_0},
\]
where the brane conformal time $t_\pm$ is defined on the branes via $n\d
t = b\d t_\pm$.  
When augmented with the initial conditions that $n$ and
$b$ both tend to unity as $x$ tends to zero (so that we recover
compactified Milne spacetime near the collision), these two equations are fully
sufficient to determine the two $x$-dependent moduli to all orders in $y_0$.


Putting the above into practice, for convenience we will work with
the Neumann variables $\tN$ and $q$, generalizing 
(\ref{ad_n_series}) and (\ref{ad_q}) to
\[
n^{-1} = \tN(x,\w)-\w x,  \qquad  \frac{b}{n}=q(x,\w).
\]
Seeking an expansion of the form
\bea
\label{ad_ansatz1}
\tN(x,\w) &=& \tN_0(x,\w) + y_0^2\tN_1(x,\w)+O(y_0^4) \\
\label{ad_ansatz2}
q(x,\w) &=& q_0(x,\w)+y_0^2\, q_1(x,\w)+O(y_0^4),
\eea
the Einstein equations (\ref{e1}) and (\ref{e2}) when expanded to zeroth order in
$y_0$ immediately restrict $\tN_0$ and $q_0$ to be functions of 
$x$ alone.  The bulk geometry to this order is then simply anti-de Sitter space with
time-varying moduli, consistent with (\ref{ad_n_series}) and (\ref{ad_q}).
The moduli $\tN_0(x)$ and $q_0(x)$ may be found
by integrating the brane equations (\ref{G55}), also expanded to lowest order in $y_0$.
In terms of the Lambert W-function \cite{LambertW}, $W(x)$, defined implicitly by
\[
\label{Wdef}
W(x)e^{W(x)}=x,
\]
the solution is
\[
\label{sol1}
\tN_0(x) = e^{\frac{1}{2}W(-x^2)}, \qquad q_0(x) = \left(1+W(-x^2)\right)\,e^{\frac{1}{2}W(-x^2)}.
\]
Thus we have found the scaling solution for the background, whose form is
independent of $y_0$, holding for any $y_0\ll 1$.
Using the series expansion for the Lambert W-function about
$x=W(x)=0$, namely \cite{comment1}
\[
\label{W_series}
W(x) = \sum_{m=1}^\inf\frac{(-m)^{m-1}}{m!}x^m,
\]
we can immediately check that the expansion of our solution is in
exact agreement with (\ref{N_series}) and (\ref{q_series}). 
At leading order in $y_0$ then, we have succeeded in resumming  
the Dirichlet/Neumann 
polynomial expansion results for the background to all orders in $x$.

Later, we will return to evaluate the $y_0^2$ corrections in our
expansion about the scaling solution.  
In the next two subsections, however, we will first examine the
scaling solution in greater detail. 

\subsection{Evolution of the brane scale factors}

Using the scaling solution (\ref{sol1}) to evaluate the scale factors on both
branes, we find to $O(y_0^2)$
\[
b_\pm = 1\pm x e^{-\frac{1}{2}W(-x^2)} = 1\pm\sqrt{-W(-x^2)}.
\]
To follow the evolution of the brane scale factors, it is helpful to
first understand the behavior of the Lambert W-function, the real
values of which are displayed in Figure 2. 
\begin{figure}
\label{Wfigure}
\includegraphics[width=12cm]{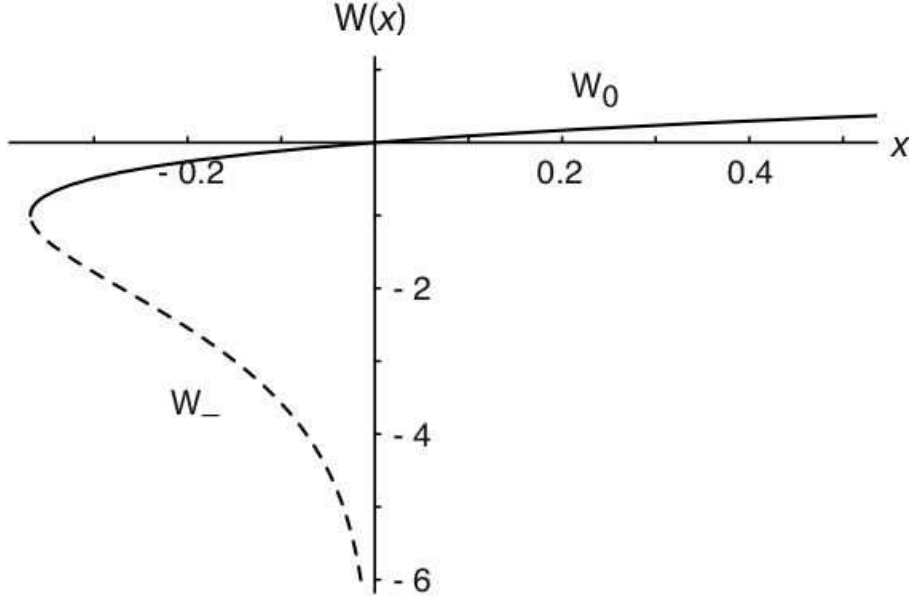}
\caption{
The real values of the Lambert W-function.  The solid line indicates
the principal solution branch, $W_0(x)$, while the dashed line depicts the
$W_{-1}(x)$ branch.  The two branches join smoothly at $x=-1/e$ where 
$W$ attains its negative maximum of $-1$.}
\end{figure}
For positive arguments the Lambert W-function is single-valued,
however, for the negative arguments of interest here, we see that there
are in fact two different real solution branches.
The first branch, denoted $W_0(x)$, satisfies $W_0(x)\ge -1$ and is usually referred
to as the principal branch, while the second branch, $W_{-1}(x)$,
is defined in the range $W_{-1}(x)\le -1$.  The two solution branches
join smoothly at $x=-1/e$, where $W=-1$. 

Starting at the brane collision where $x=0$, the brane scale factors
are chosen to satisfy $b_\pm=1$, and so we must begin on the
principal branch of the Lambert W-function 
for which $W_0(0)=0$.  Thereafter, as illustrated in Figure 3, 
$b_+$ increases and $b_-$ decreases monotonically until 
at the critical time $x=x_c$, when $W_0(-x_c^2)=-1$ and $b_-$ shrinks
to zero.  From (\ref{Wdef}), the critical time is therefore
$
x_c = e^{-\frac{1}{2}} = 0.606...,
$
and corresponds physically to the time at which the negative-tension brane
encounters the bulk black hole \cite{comment2}.

At this moment, the scale factor on the positive-tension brane 
has only attained a value of two.  
From the Birkhoff-frame solution, in which the bulk is
Schwarzschild-AdS and the branes are moving, we know that the
positive-tension brane is unaffected by the disappearance 
of the negative-tension brane and simply continues its journey out to the boundary of AdS. 
To reconcile this behavior with our solution in brane-static
coordinates, it is helpful to pass to $t_+$, the conformal time on the
positive-tension brane.  Working to zeroth order in $y_0$, this
may be converted into the dimensionless form
\[
\label{x4}
x_4= \frac{y_0 t_+}{L} =\frac{y_0}{L}\int \frac{n}{b}\,\d t= \int \frac{\d x}{q_0(x)}=
x e^{-\frac{1}{2}W(-x^2)}=\sqrt{-W(-x^2)} .
\]
Inverting this expression, we find that the bulk time parameter $x=x_4\,e^{-\frac{1}{2}x_4^2}$.
The bulk time $x$ is thus double-valued when expressed as a function
of $x_4$, the conformal time on the positive-tension brane: to continue
forward in $x_4$ beyond $x_4=1$ (where $x=x_c$), the bulk time $x$ must
reverse direction and decrease towards zero.  The metric functions,
expressed in terms of $x$, must then continue back along the other branch
of the Lambert W-function, namely the $W_{-1}$ branch.
In this manner we see that the solution for the scale factor on the positive-tension
brane, when continued on to the $W_{-1}$ branch, tends to infinity as
the bulk time $x$ is reduced back towards zero (see dotted line in Figure 3),
corresponding to the positive-tension brane approaching the boundary
of AdS as $x_4 \tt \inf$.

For simplicity, in the remainder of this paper we will work
directly with the brane conformal time $x_4$ itself.  With this choice,
the brane scale factors to zeroth order in $y_0$ are simply
\nopagebreak[1] $b_\pm = 1\pm x_4$.

\begin{figure}
\label{bfigure}
\includegraphics[width=11cm]{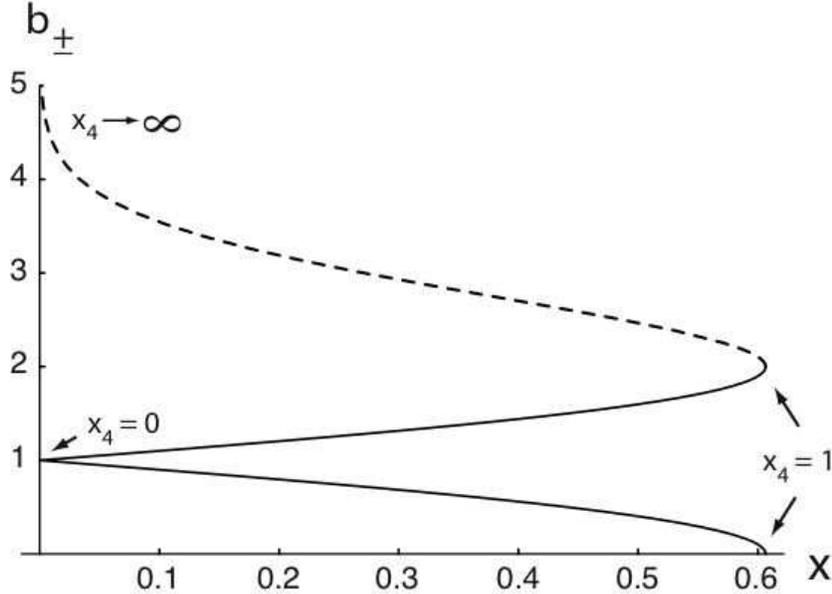} 
\caption{
 The scale factors $b_{\pm}$ on the positive-tension brane (rising curve)
and negative-tension brane (falling curve) as a function of the bulk time
 parameter $x$, to zeroth order in $y_0$.  The continuation of the
 positive-tension brane scale factor on to the $W_{-1}$ branch 
of the Lambert W-function is indicated by the dashed line.  }
\end{figure}

\subsection{Analytic continuation of the bulk geometry}

In terms of $x_4$, the metric functions $n$ and $b$ are given by
\[
\label{nandb}
n = \frac{e^{\frac{1}{2}x_4^2}}{1-\w x_4}+O(y_0^2), \qquad b = \frac{1-x_4^2}{1-\w x_4}+O(y_0^2).
\]
At $x_4=1$, the three-dimensional scale factor $b$ shrinks to zero at
all values of $\w$ except $\w=1$ (\ie the positive-tension brane).
Since $b$ is a coordinate scalar under transformations of $x_4$ and
$\w$, one might be concerned that that the scaling solution becomes singular at
this point.  However, when we compute the $y_0^2$ corrections as we
will do shortly, we will 
find that the corrections become large close
to $x_4=1$, precipitating a breakdown of the small-$y_0$
expansion.  
Since it will later turn out that the 
scaling solution
maps directly on to the four-dimensional
effective theory, and that this, like the metric on the
positive-tension brane, is completely regular at $x_4=1$, we are
encouraged to simply analytically continue the scaling solution to times
$x_4>1$.  

When implementing this analytic continuation careful attention must be paid to
the range of the coordinate $\w$.  Thus far, for times $x_4<1$, we
have regarded $\w$ as a coordinate spanning the fifth dimension,
taking values in the range $-1\le \w \le 1$.  The two metric functions
$n$ and $b$ were then expressed in terms of the coordinates $x_4$ and $\w$.
Strictly speaking, however, this parameterization is redundant: we could
have chosen to eliminate $\w$ by promoting the three-dimensional scale factor $b$ 
from a metric function to an independent coordinate parameterizing the
fifth dimension.  Thus we would have only one metric function $n$,
expressed in terms of the coordinates $x_4$ and $b$.   

While this latter parameterization is more succinct, its disadvantage is
that the locations of the branes are no longer explicit, since the
value of the scale factor $b$ on the branes is time-dependent.
In fact, to track the location of the branes we must re-introduce the
function $\w(x_4,b)=(b+x_4^2-1)/bx_4$ (inverting (\ref{nandb}) at
lowest order in $y_0$).  The trajectories of the branes are then 
given by the contours $\w=\pm 1$.

\begin{figure}
\label{coathanger}
\includegraphics[width=12cm]{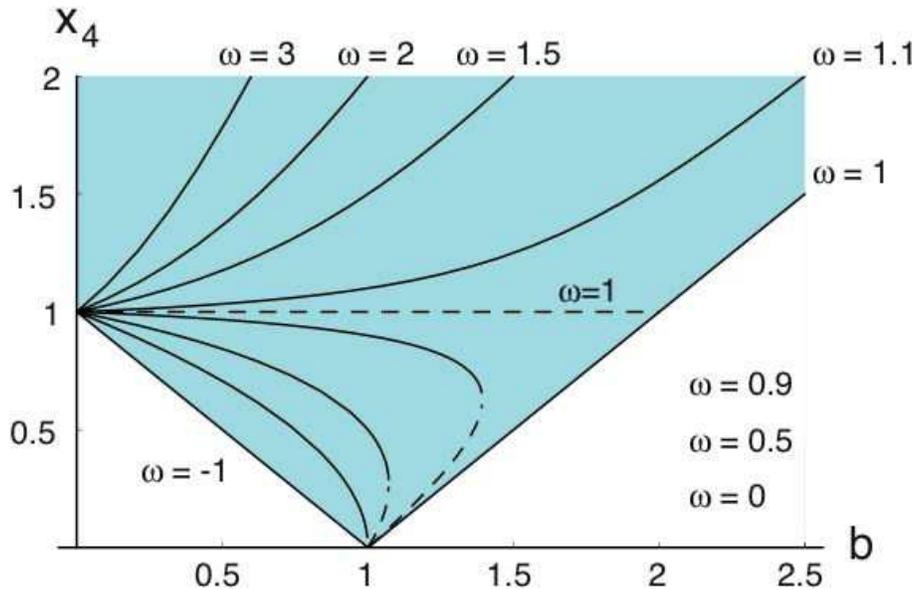} 
\caption{
The contours of constant $\w$ in the ($b$, $x_4$) plane.
Working to zeroth order in $y_0$, these are given by $x_4 =
\frac{1}{2}\big(b\w\pm \sqrt{b^2\w^2-4(b-1)}\big)$, where we have
plotted the positive root using a solid line and the negative root using a dashed line.
The negative-tension brane is located at $\w=-1$ for times $x_4<1$,
and the trajectory of the positive-tension brane is given (for all time)
by the positive root solution for $\w=1$.  The region delimited by the
trajectories of the branes (shaded) then corresponds to the bulk.
From the plot we see that, for $0<x_4<1$, the bulk is parameterized by
values of $\w$ in the range $-1\le \w \le 1$.  In contrast, for $x_4>1$, the
bulk is parameterized by values of $\w$ in the range $\w\ge 1$.
} 
\end{figure}

The contours of constant $\w$ as a function of $x_4$ and $b$ are
plotted in Figure 4.  The analytic continuation to times $x_4>1$ has
been implemented, and the extent of the bulk is indicated by
the shaded region.
From the figure, we see that, if we were to revert to our original
parameterization of the bulk in terms of $x_4$ and $\w$, the range of
$\w$ required depends on the time coordinate $x_4$: for early times
$x_4<1$, we require only values of $\w$ in the range $-1\le \w \le 1$,
whereas for late times $x_4>1$, we require values in the range $\w\ge
1$.  Thus, while the positive-tension brane remains fixed at $\w=1$
throughout, at early times $x_4<1$ the value of $\w$ {\it decreases} as we
head away from the positive-tension brane along the fifth dimension, whereas
at late times $x_4>1$, the value of $\w$ {\it increases} away from the positive-tension brane.

While this behavior initially appears paradoxical if $\w$ is
regarded as a coordinate along the fifth dimension, we stress that 
the only variables with meaningful physical content are the brane conformal time $x_4$ and the
three-dimensional scale factor $b$.  These physical variables behave
sensibly under analytic continuation.   
In contrast, $\w$ is simply a convenient
parameterization introduced to follow the brane trajectories, with the
awkward feature that its range alters under the analytic continuation at $x_4=1$.

\begin{figure}
\label{bplots}
\begin{tabular}{cc}
\begin{minipage}{8cm}
\hspace{-0.8cm}
\includegraphics[width=8cm]{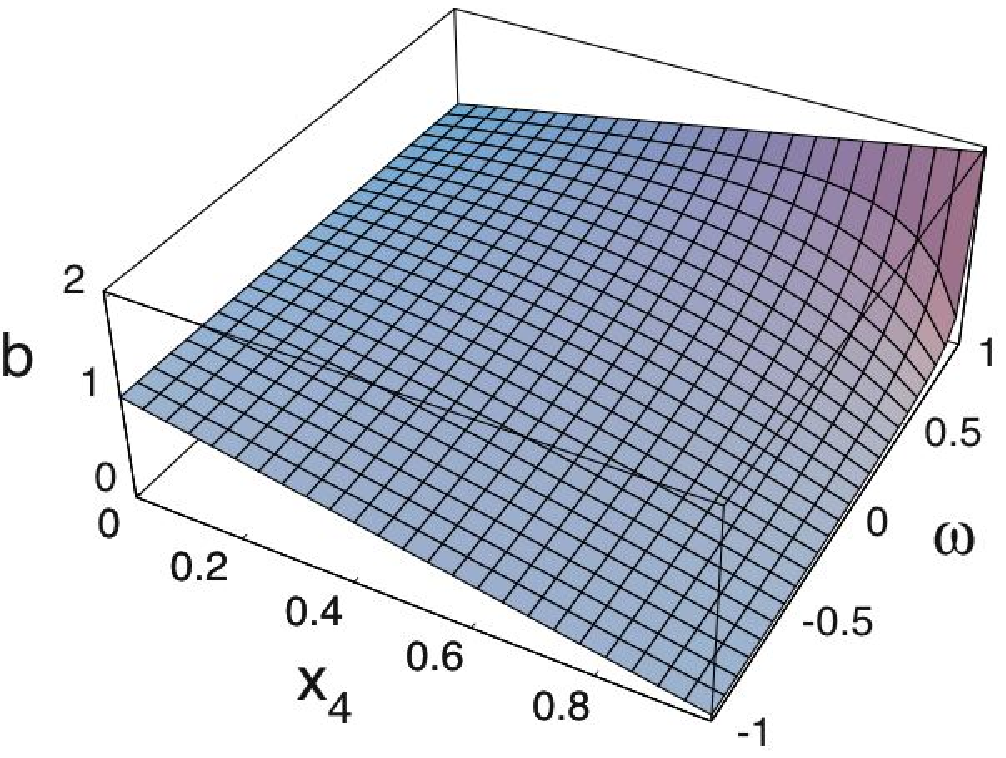}
\end{minipage}
\hspace{0cm}
\begin{minipage}{8cm}
\includegraphics[width=8cm]{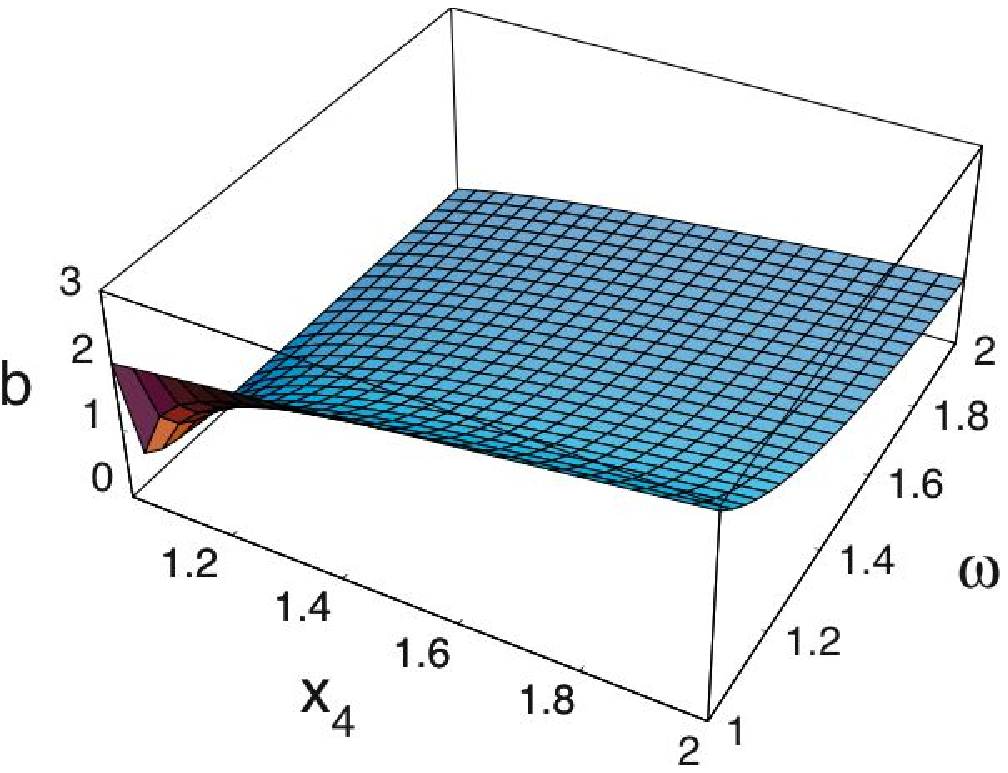}
\end{minipage}
\end{tabular}
\caption{
The three-dimensional scale factor $b$, plotted to zeroth order in $y_0$ as a
function of $x_4$ and $\w$, for $x_4<1$ (left) and $x_4>1$
(right). The positive-tension brane is fixed at $\w=1$ for all time
(note the evolution of its scale factor is smooth and continuous),
and for $x_4<1$, the negative-tension brane is located at $\w=-1$.  
}
\end{figure}

For the rest of this paper, we will find it easiest to continue parameterizing the bulk
in terms of $x_4$ and $\w$, adjusting the range of the $\w$ where required.
Figure 5 illustrates this approach: at early times $x_4<1$ the
three-dimensional scale factor $b$ is plotted for values of $\w$ in
the range $-1\le \w \le 1$.  At late times $x_4>1$, we must however
plot $b$ for values of $\w$ in the range $\w\ge 1$.  In this fashion,
the three-dimensional scale factor $b$ always decreases along the
fifth dimension away from the brane.  

We have argued that 
the scaling solution for the background, obtained
at lowest order in $y_0$, may 
be analytically continued across $x_4=1$. 
There is a coordinate singularity in the $x_4$, $\w$ 
coordinates but this does not affect the metric on the 
positive-tension brane which remains regular throughout.
The same features will be true when we solve for the
cosmological perturbations. The fact that the
continuation is regular on the positive-tension brane
and precisely agrees with the predictions of the
four-dimensional effective theory provides strong evidence for its
correctness. Once the form of the 
the background and the perturbations have been 
determined to lowest order in $y_0$, the 
higher-order corrections are obtained from
differential equations in $y$ with source
terms depending only on the lowest order solutions. 
It is straightforward to obtain these corrections
for $x_4<1$. If we analytically continue them as
described to $x_4>1$ as described, 
we automatically solve the bulk
Einstein equations and the Israel matching 
conditions on the positive
tension brane for all $x_4$. The continued
solution is well-behaved after the collision 
in the vicinity of the 
positive-tension brane, out to large distances where
the $y_0$ expansion fails. 
%

\subsection{
Higher-order corrections}

In this section we explicitly compute the 
$y_0^2$ corrections.  The size of these
corrections indicates the validity of the expansion about the scaling solution, which
perforce is only valid when the $y_0^2$ corrections are small.

Following the procedure outlined previously, we first evaluate
the Einstein equations (\ref{e1}) and (\ref{e2}) to $O(y_0^2)$ using the ans{\"a}tze
(\ref{ad_ansatz1}) and (\ref{ad_ansatz2}), along with the solutions for
$\tN_0(x)$ and $q_0(x)$ given in (\ref{sol1}). 
The result is two second-order ordinary differential equations in
$\w$, which may straightforwardly be integrated yielding
$\tN_1(x,\w)$ and $q_1(x,\w)$ up to two arbitrary functions of $x_4$. 
These time-dependent moduli are then fixed using the brane equations (\ref{G55}), evaluated
at $O(y_0^2)$ higher than previously.  

We obtain the result:
\bea
\label{ad_n}
n(x_4,\w) &=& \frac{e^{\frac{1}{2}x_4^2}}{1 - \w x_4} + \frac{e^{\frac{1}{2}x_4^2}
  y_0^2}{30 {\left( -1 + \w x_4 \right) }^2 {\left( -1 + x_4^2 \right)
  }^4} \nonumber \\  
    && \big( x_4 \big( 5 \w \left( -3 + \w^2 \right)  - 5 x_4 +
     40 \w \left( -3 + \w^2 \right)  x_4^2 -  
          5 \left( -14 + 9 \w^2 \left( -2 + \w^2 \right)  \right)
	   x_4^3 \nonumber \\ && + 3 \w^3 \left( -5 + 3 \w^2 \right)  x_4^4 - 19 x_4^5 +  
          5 x_4^7 \big)  - 5 {\left( -1 + x_4^2 \right) }^3 \ln (1 -
	  x_4^2) \big) +O(y_0^4), \\
&&\nonumber \\
\label{ad_b}
b(x_4,\w) &=& 
\frac{1 - x_4^2}{1 - \w x_4} + \frac{x_4 y_0^2}{30  
     {\left( -1 + \w x_4 \right) }^2 {\left( -1 + x_4^2 \right) }^3}
\nonumber \\ && \big( -5 \w \left( -3 +
  \w^2 \right)  - 20 x_4 + 5 \w \left( -7 + 4 \w^2 \right)  x_4^2 -  
       10 \left( 1 - 12 \w^2 + 3 \w^4 \right)  x_4^3 \nonumber \\ & & + 3 \w \left(
       -20 - 5 \w^2 + 2 \w^4 \right)  x_4^4 - 12 x_4^5 + 31 \w x_4^6 -  
       5 \w x_4^8 \nonumber \\ && - 5 {\left( -1 + x_4^2 \right) }^2 \left( \w - 2 x_4 +
       \w x_4^2 \right)  \ln (1 - x_4^2) \big) +O(y_0^4).
\eea

\begin{figure}
\label{metricplots}
\begin{tabular}{cc}
\begin{minipage}{8cm}
\hspace{-0.8cm}
\includegraphics[width=8cm]{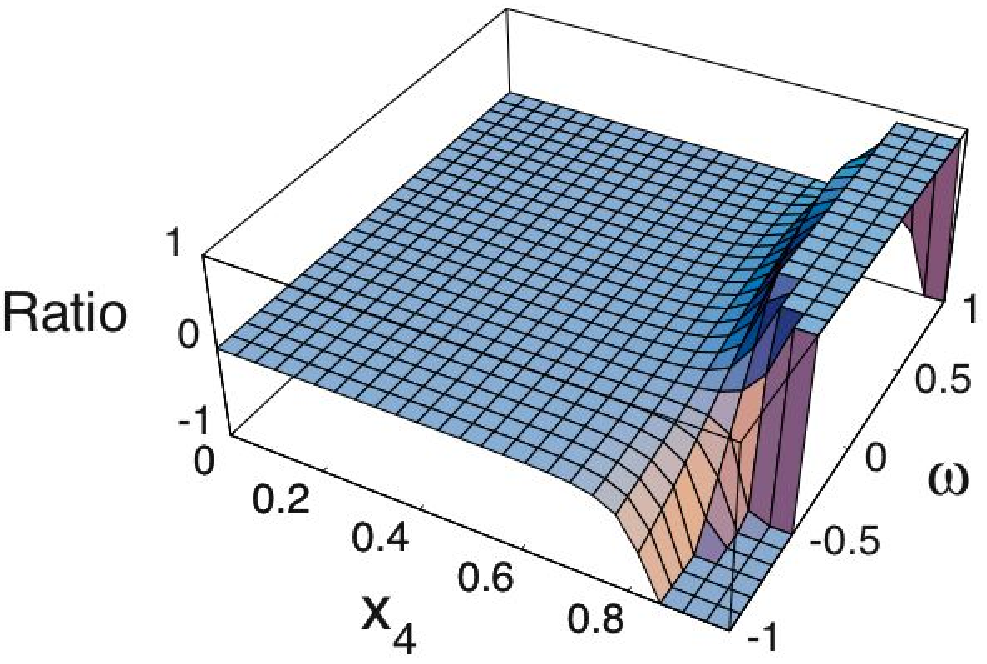}
\end{minipage}
\hspace{0cm}
\begin{minipage}{8cm}
\includegraphics[width=8cm]{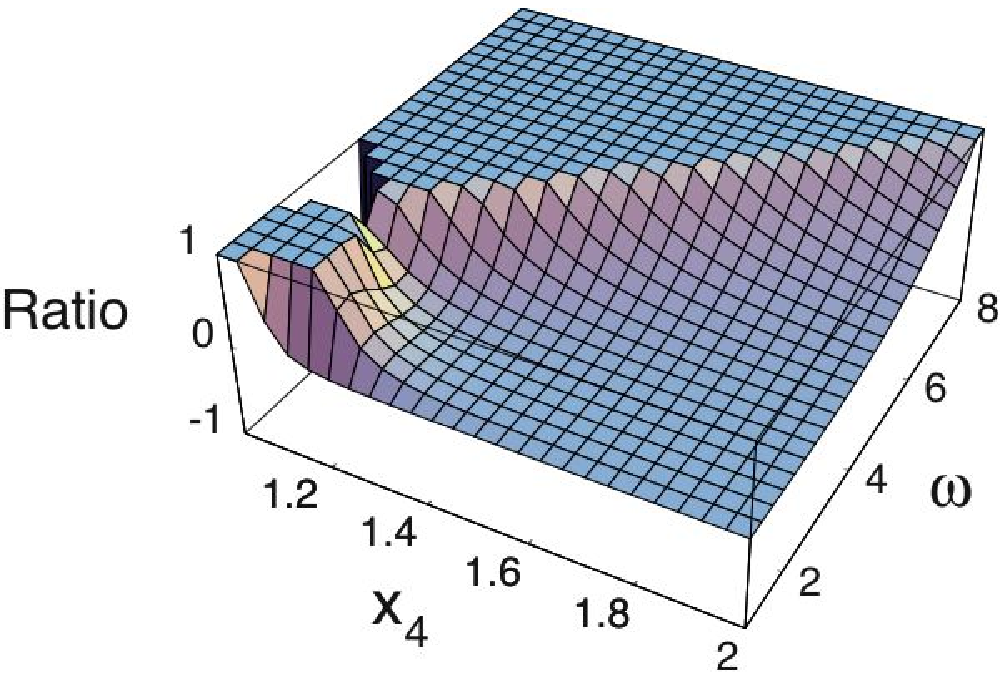}
\end{minipage}
\end{tabular}
\caption{
The ratio of the $y_0^2$ corrections to the leading term in the
small-$y_0$ expansion for $b$, plotted for $x_4<1$ (left) and $x_4>1$
(right), for the case where $y_0=0.1$.   Where this ratio becomes of
order unity the expansion about the scaling solution breaks down.  
The analogous plots for $n$ display similar behavior.
}
\end{figure}
In Figure 6, we have plotted the ratio of the $y_0^2$ corrections to
the corresponding terms at leading order: where this ratio becomes of
order unity the expansion about the scaling solution breaks down.  
Inspection shows there are two such regions: the
first is for times close to $x_4=1$, for all $\w$, and the second
occurs at late times $x_4>1$, far away from the positive-tension brane.
In neither case does the failure of the $y_0$ expansion indicate a
singularity of the background metric: from the bulk-static coordinate
system we know the exact solution for the background metric is simply
Schwarzschild-AdS, which is regular everywhere.  
The
exact bulk-static solution in Birkhoff frame  
tells us that the proper speed of the negative-tension brane,
relative to the static bulk, approaches the speed of light 
as it reaches the event horizon
of the bulk black hole.  It therefore seems plausible 
that a small-$y_0$ expansion 
based upon slowly-moving branes must break
down at this moment, when $x_4=1$ in our chosen
coordinate system.

Analytically continuing our solution in both $x_4$ and $\w$ around
$x_4=1$, the logarithmic terms in the $y_0^2$ corrections now
acquire imaginary pieces for $x_4>1$.  However, since these
imaginary terms are all suppressed by a factor of $y_0^2$, they
can only enter the Einstein-brane equations (expanded as a series
to $y_0^2$ order) in a linear fashion.
Hence the real and imaginary parts of the metric necessarily
constitute {\it independent} solutions, permitting us to simply
throw away the imaginary part and work with the real part alone.
As a confirmation of this, it can be checked explicitly that replacing
the $\ln{(1-x_4)}$ terms in (\ref{ad_n}) and (\ref{ad_b}) with
$\ln{|1-x_4|}$ still provides a valid solution to $O(y_0^2)$ of the complete
Einstein-brane equations and boundary conditions.  

Finally, at late times where $x_4>1$, note that the extent to which we know the
bulk geometry away from the positive-tension brane is limited by the
$y_0^2$ corrections, which become large at an increasingly large value
of $\w$, away from the positive-tension brane (see Figure 6).
The expansion about the scaling solution thus breaks down before we
reach the horizon of the bulk black hole, which is located at $\w\tt \inf$ for $x_4>1$.

\subsection{Treatment of the perturbations}

Having determined the background geometry to $O(y_0^2)$ in the
preceding subsections, we now turn our attention to the perturbations.
In this subsection we show how to evaluate the perturbations to
$O(y_0^2)$ by expanding about the scaling solution.  The results will
enable us to perform stringent checks of the four-dimensional
effective theory and moreover to evaluate the mode-mixing between early and
late times.

In addition to the dimensionless variables $x=y_0 ct/L$ and $\w = y/
y_0$, when we consider the metric perturbations we must further introduce 
the dimensionless perturbation amplitude $\tB = B y_0^2 c^2/L^2 \sim B
V^2/L^2$ and the
dimensionless wavevector $\tk = kL/y_0\sim ckL/V$.  In this fashion, 
to lowest order in $y_0$ and $k$, we then find $\Phi_L = A - B/t^2
= A - \tB /x^2$ and similarly $kct=\tk x$.  (Note that
the perturbation amplitude A is already dimensionless however).

Following the treatment of the perturbations in the Dirichlet/Neumann 
polynomial expansion,
we will again express the metric perturbations in terms of $W_L$
(obeying Dirichlet boundary conditions), and the Neumann variables
$\phi_4$ and $\xi_4$, defined in (\ref{phi4xi4}).  
Hence we seek an expansion of the form
\bea
\label{adperts}
\phi_4(x_4,\w)&=&\phi_{40}(x_4,\w)+y_0^2\,\phi_{41}(x_4,\w)+O(y_0^4), \\
\xi_4(x_4,\w) &=& \xi_{40}(x_4,\w)+y_0^2\,\xi_{41}(x_4,\w)+O(y_0^4),\\
W_L(x_4,\w)&=& W_{L0}(x_4,\w)+y_0^2\,W_{L1}(x_4,\w)+O(y_0^4).
\eea

As in the case of the background, we will use the $G^5_5$ equation
evaluated on the brane to fix the arbitrary functions of
$x_4$ arising from integration of the Einstein equations with respect to
$\w$.  By substituting
the Israel matching conditions into the $G^5_5$ equation, along with
the boundary conditions for the perturbations, it is possible to 
remove the single $\w$-derivatives that appear.  
We arrive at the following 
second-order ordinary differential equation, valid on both branes, 
\bea
\label{breq}
0&=&
 2n(x_4^2-1)(2b^2\phi_4+\xi_4)\dot{b}^2+b^2\left(nx_4(x_4^2-3)-(x_4^2-1)\dot{n}\right)
(2b^2\dot{\phi}_4-\dot{\xi}_4)\nonumber 
\\ &&
+b\dot{b}\left(4nx_4(x_4^2-3)(b^2\phi_4+\xi_4)-(x_4^2-1)(4(b^2\phi_4+\xi_4)\dot{n}
-n(10b^2\dot{\phi}_4+\dot{\xi}_4))\right)\nonumber 
\\ &&+bn(x_4^2-1)\left(4(b^2\phi_4+\xi_4)\ddot{b}+2b^3\ddot{\phi}_4-b\ddot{\xi}_4\right),
\eea
where dots indicate differentiation with respect to $x_4$, 
and where, in the interests of brevity, we have omitted terms of $O(\tk^2)$. 

Beginning our computation, the $G^5_i$ and $G^5_5$ Einstein  
equations when evaluated to lowest order in $y_0$ immediately restrict
$\phi_{40}$ and $\xi_{40}$ to be functions of $x_4$ only.
Integrating the $G^0_i$ equation with respect to $\w$ then gives $W_{L0}$
in terms of $\phi_{40}$ and $\xi_{40}$, up to an arbitrary function of
$x_4$.  Requiring that $W_{L0}$ vanishes on both branes allows us to
both fix this arbitrary function, and also to solve for $\xi_{40}$ in
terms of $\phi_{40}$ alone.  Finally, evaluating (\ref{breq}) on both branes 
to lowest order in $y_0$ and solving simultaneously yields a
second-order ordinary differential equation for $\phi_{40}$, with
solution 
\[
\label{phi40soln}
\phi_{40} =
\left(\frac{3A}{2}-\frac{9\tB}{16}\right)-\frac{\tB}{2x_4^2} +
O(\tk^2),  
\]
where the two arbitrary constants have been chosen to match the
small-$t$ series expansion given in Section III.
With this choice,
\bea
\label{xi4soln}
\xi_{40} &=&
-A+\frac{11\tB}{8}-\frac{\tB}{x_4^2}+\left(A-\frac{3\tB}{8}\right)x_4^2
+O(\tk^2)\\ 
W_{L0} &=& e^{-\frac{1}{2}x_4^2}\frac{(1-\w^2)}{(1-x_4^2)^2}\left(3
Ax_4^2(-2+\w 
  x_4)+\tB x_4(\frac{9}{4}x_4+\w (1-\frac{9}{8}x_4^2))\right) +O(\tk^2).
\eea
The resulting behavior for the perturbation to the three-dimensional
scale factor, $b^2 \Psi_L$, is plotted in Figure 7.

\begin{figure}
\label{pertplots}
\begin{tabular}{cc}
\begin{minipage}{8cm}
\hspace{-0.8cm}
\includegraphics[width=8cm]{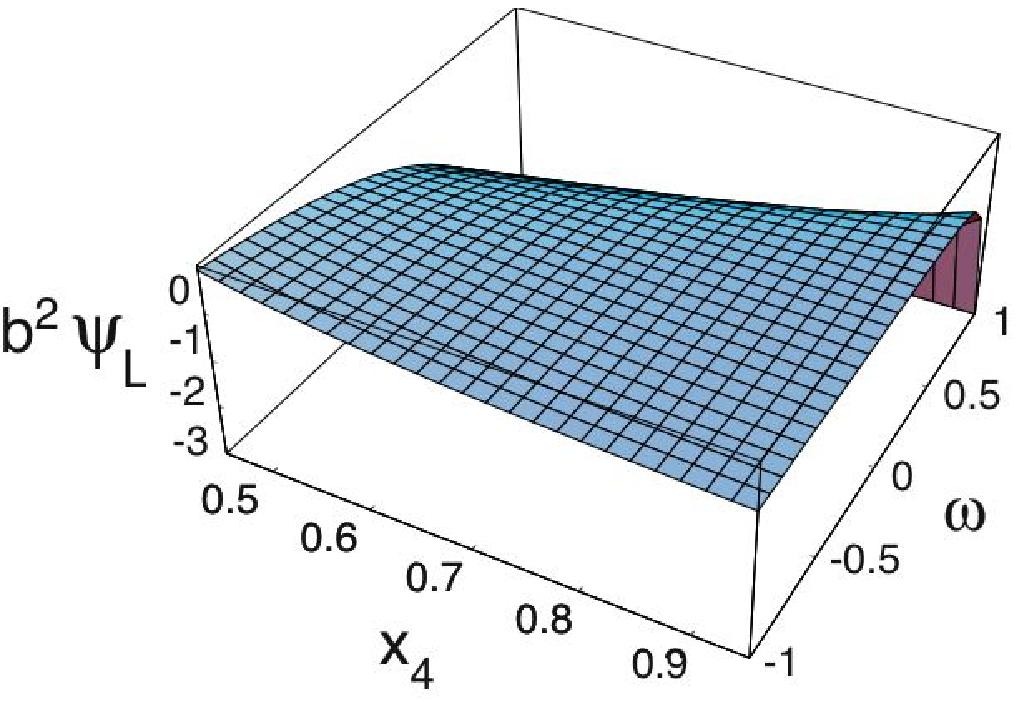}
\end{minipage}
\hspace{0cm}
\begin{minipage}{8cm}
\includegraphics[width=8cm]{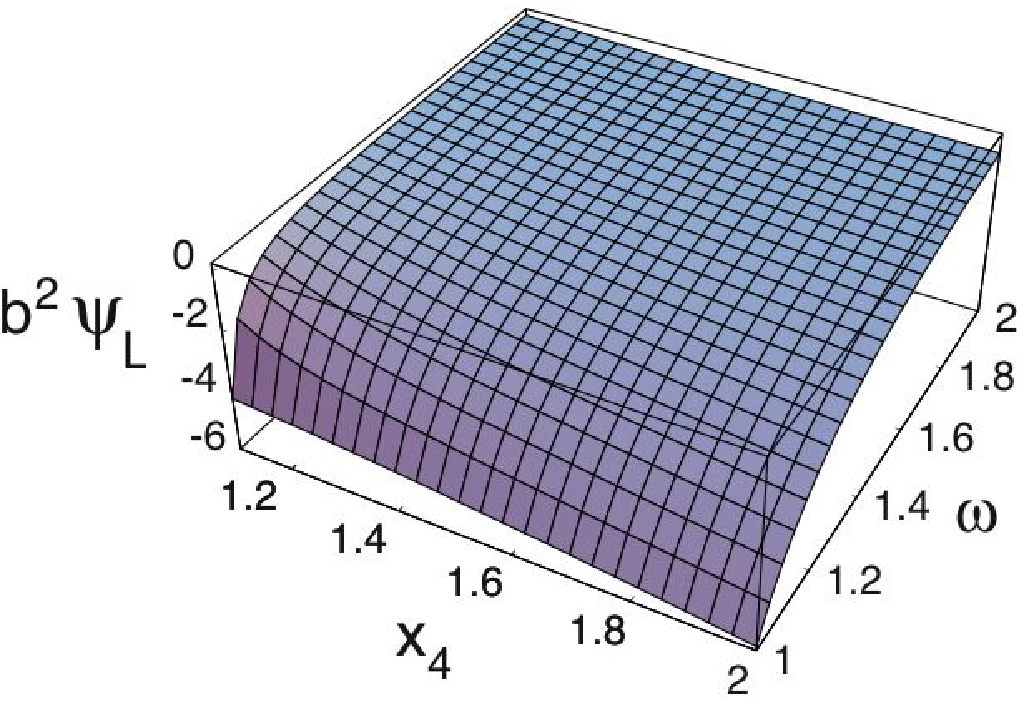}
\end{minipage}
\end{tabular}
\caption{
The perturbation to the three-dimensional scale factor, $b^2 \Psi_L$, plotted on long
wavelengths to zeroth order in $y_0$ for early times (left) and late times (right).
Only the $\tB$ mode is displayed (\ie $A=0$ and $\tB=1$).  Note
how the perturbations are localized on the positive-tension brane
(located at $\w =1$), and decay away from the brane.
}
\end{figure}
In terms of the original Newtonian gauge variables, an 
identical calculation (working now to all orders in $\tk$) yields,
{\allowdisplaybreaks
\bea
\label{PhiLallk}
\Phi_L &=& \frac{2\,\tk \,(1-\w x_4)^2}{3\, (x_4^2-1)}\,\big(A_0 J_0(\tk
x_4)+B_0 Y_0(\tk x_4)\big)\nonumber \\ && \qquad +\frac{1}{x_4}\left(1+\frac{(1-\w
  x_4)^2}{1-x_4^2}\right)\big(A_0 J_1(\tk x_4)+B_0 Y_1(\tk x_4)\big)+O(y_0^2),
\\
\label{PsiLallk}
\Psi_L &=& \frac{1}{3\,(1-x_4^2)}\Big(2\tk (1-\w x_4)^2 \big(A_0 J_0(\tk
x_4)+B_0 Y_0(\tk x_4)\big) \nonumber \\ && \qquad - 3\,(x_4+\w(-2+\w x_4))\big(A_0 J_1(\tk
x_4)+B_0 Y_1(\tk x_4)\big)\Big) +O(y_0^2), \\
 W_{L} &=&
2\,x_4^2
\,e^{-\frac{1}{2}x_4^2}\,\frac{(\w^2-1)}{(1-x_4^2)^2}\,\Big(\tk\, (1-\w
 x_4)\big(A_0 J_0(\tk x_4)+B_0
 Y_0(\tk x_4)\big)\nonumber \\
 &&\qquad  +\w\,\big(A_0 J_1(\tk
 x_4)+B_0 Y_1(\tk x_4)\big) \Big) + O(y_0^2),
\eea
}
where the constants $A_0$ and $B_0$ are given by
\[
\label{A0B0}
A_0 = \frac{3A}{\tk}-\frac{9\tB}{8\tk}+\frac{1}{2}\tB \tk (\ln{2}-\gamma)+O(y_0^2), \qquad B_0 =
  \frac{\tB \tk \pi}{4} +O(y_0^2).
\]

To evaluate the $y_0^2$ corrections, we repeat the same sequence of
steps: integrating the $G^5_i$ and $G^5_5$ Einstein equations (at
$y_0^2$ higher order) gives us $\phi_{41}$ and $\xi_{41}$ up to two
arbitrary functions of $x_4$, and integrating the $G^0_i$ equation
then gives us $W_{L1}$ in terms of these two arbitrary functions plus
one more.   Two of the three arbitrary functions are then determined
by imposing the Dirichlet boundary conditions on $W_{L1}$, and the
third is found to satisfy a second-order ordinary differential
equation after making use of (\ref{breq}) on both branes.
This can be solved, and the constants of integration appearing in the solution are again chosen
so as to match the small-$t$ series expansion of Section III.

Converting back to the original longitudinal gauge variables, the
results to $O(y_0^4)$ and to $O(k^2)$ take the schematic form 
\bea
\Phi_L &=&  f^\Phi_0+y_0^2 (f^\Phi_1+f^\Phi_2
\ln{(1+x_4)} + f^\Phi_3 \ln{(1-x_4)}+f^\phi_4 \ln{(1-\w x_4)}, \\
\Psi_L &=&  f^\Psi_0+y_0^2 (f^\Psi_1+f^\Psi_2
\ln{(1+x_4)} + f^\Psi_3 \ln{(1-x_4)}+f^\Psi_4 \ln{(1-\w x_4)}, \\
W_L&=& e^{-\frac{1}{2}x_4^2}\left( f^W_0+y_0^2 (f^W_1+f^W_2
\ln{(1+x_4)} + f^W_3 \ln{(1-x_4)}+f^W_4 \ln{(1-\w x_4)})\right),
\eea
where the $f$ are rational functions of $x_4$ and $\w$ which, due to
their length, have been listed separated separately in Appendix D.
(If desired, more detailed results including the $O(k^2)$ corrections
are available \cite{Website}). 

It is easy to check that the results obtained by expanding about the
scaling solution are consistent with those obtained using our previous
method based upon Dirichlet/Neumann 
polynomials.  Taking the results
from the polynomial expansion given in Appendix C, 
substituting $t=(x_4/y_0) e^{-\frac{1}{2}x_4^2}$ and $y=\w y_0$,
retaining only terms of $O(y_0^2)$ or less, one finds agreement with
the results listed in Appendix D after these have been re-expressed as a
series in $x_4$.  This has been checked explicitly both for the
background and the perturbations.

Just as in the case of the background, the small-$y_0$ expansion breaks down
for times close to $x_4=1$ when the $y_0^2$ corrections to the
perturbations become larger than the corresponding zeroth order terms.
Again, we will simply analytically continue the solution in $x_4$ and
$\w$ around this point.  In support of this, the induced metric on the
positive-tension brane is, to zeroth order in $y_0$, completely
regular across $x_4=1$, even including the perturbations as
can be seen from (\ref{PhiLallk}) and (\ref{PsiLallk}).

As in the case of the background, any imaginary pieces acquired from
analytically continuing logarithmic terms are all suppressed by order
$y_0^2$.  Thus they may only enter the Einstein-brane equations
(when these are expanded to order $y_0^2$) in a linear fashion, and
hence the real and imaginary parts of the metric constitute independent
solutions.  We can therefore simply drop the imaginary parts, or
equivalently replace the $\ln{(1-x_4)}$ and $\ln{(1-\w x_4)}$ terms
with $\ln{|1-x_4|}$ and $\ln{|1-\w x_4|}$ respectively.  
We have checked explicitly that this still satisfies the Einstein-brane
equations and boundary conditions.

\section{Comparison with the Four-dimensional Effective Theory}

We have now arrived at a vantage point from which we may scrutinize
the predictions of the four-dimensional effective theory in light of our
expansion of the bulk geometry about the scaling solution.
We will find that the four-dimensional effective theory is in exact
agreement with the scaling solution.
Beyond this, the $y_0^2$ corrections lead to effects 
that cannot be described within a four-dimensional effective framework.
Nonetheless, the higher-order corrections are automatically small at
very early and very late times, restoring the accuracy of the
four-dimensional effective theory in these limits.

In the near-static limit, the mapping from four to five dimensions
may be calculated from the moduli space approach \cite{Terning, 
Ekpyrotic,KhouryZ}: putting the four-dimensional effective theory metric
$g^4_{\mu\nu}$ into Einstein frame, the mapping
reads
\[
g_{\mu \nu}^{+}= \cosh^2 (\phi/\sqrt{6})g_{\mu \nu}^{4} \qquad 
g_{\mu \nu}^{-}= \sinh^2 (\phi/\sqrt{6})g_{\mu \nu}^{4},
\label{map4}
\]
where $g_{\mu \nu}^{+}$ and $g_{\mu \nu}^{-}$ are the metrics
on the positive- and negative-tension branes respectively, 
and $\phi$ is the radion.
Two of us have shown elsewhere that on 
symmetry grounds this is the unique local mapping
involving no derivatives \cite{Conf_sym}, and that to leading order
the action for $g_{\mu \nu}^{4}$ and $\phi$ is that for
Einstein gravity with a 
minimally coupled 
scalar field. 

Solving the four-dimensional effective theory is trivial: 
the background is conformally flat, $g^{4}_{\mu \nu}
= b_4^2(t_4)\,\eta_{\mu \nu}$, and the Einstein-scalar
equations yield the following solution, unique up to a 
sign choice for
$\phi_0$, of the form
\[
b_4^2=\bar{C}_4 t_4, \qquad e^{\sqrt{2\over 3}\phi_0} = \bar{A}_4 t_4,
\label{back4i}
\]
with $\phi_0$ the background scalar field, and
$\bar{A}_4$ and $\bar{C}_4$ arbitrary constants. 
(Throughout this paper we adopt units where $8 \pi G_4 =1$).

According to the map (\ref{map4}), the brane scale factors
are then predicted to be 
\[
b_\pm= {1\over 2}  b_4
e^{-{\phi_0\over \sqrt{6}}} \left( 1 \pm e^{\sqrt{2\over 3} 
\phi_0}\right) = 1\pm \bar{A}_4 t_4,
\label{back4}
\]
where we have chosen $\bar{C}_4= 4 \bar{A}_4$, so that the brane scale
factors 
are unity at the brane collision. As emphasized
in \cite{TTS}, the result (\ref{back4}) 
is actually exact for the
induced brane metrics, when $t_4$ is 
identified with the conformal time on the branes. 
From this correspondence, one can read off the five-dimensional
meaning of the 
parameter $\bar{A}_4$: it equals $L^{-1} \tanh{y_0}$ (our definition
of $y_0$ differs from that of \cite{TTS} by a factor of 2). 

With regard to the perturbations, in longitudinal gauge (see \eg 
\cite{mukhanov,bardeen}) the perturbed line element of the
four-dimensional effective theory reads
\[
\d s_4^2= b_4^2(t_4)\left[-(1+2\Phi_4)\d t_4^2+
(1-2\Phi_4)\d\vec{x}^2\right],
\label{4dmet}
\]
and the general solution to the perturbation
equations at small $k$ \cite{TTS, Khoury} is
\bea
\label{phi4eft}
\Phi_4 &=& \frac{1}{t_4}\,\left(\tA_0 J_1(k t_4)+\tB_0 Y_1(k t_4)\right),\\ 
\label{deltaradioneft}
\frac{\delta \phi}{\sqrt{6}} &=&\frac{2}{3}\,k\,\left(\tA_0
J_0(kt_4)+\tB_0Y_0(kt_4)\right) -\frac{1}{t_4}\,\left(\tA_0 J_1(kt_4)+\tB_0
Y_1(kt_4)\right),
\eea
with $\tA_0$ and $\tB_0$ being the amplitudes of the 
two linearly independent perturbation modes. 

\subsection{Background}

In the case of the background, we require 
only the result that the scale factors on the positive- and
negative-tension branes are given by
\[
b_\pm = 1\pm \bar{A}_4 t_4,
\]
where the constant $\bar{A}_4=L^{-1}\tanh{y_0}$ and $t_4$ denotes conformal
time in the four-dimensional effective theory.  (Note this solution has been
normalized so as to set the brane scale factors at the collision to unity).
Consequently, the four-dimensional effective theory restricts $b_+ + b_-=2$.  In comparison, our results from the expansion about the scaling solution
(\ref{ad_b}) give
\[
b_+ + b_- = 2 + \frac{2\,x_4^4\,\left( x_4^2-3 \right)
  \,y_0^2}{3\,{\left(1- x_4^2 \right) }^3} + O(y_0^4).
\]
Thus the four-dimensional effective theory captures the behavior of the
full theory only in the limit in which the $y_0^2$ corrections
are small, \iec when the scaling solution is an accurate description
of the higher-dimensional dynamics.  
At small times such that $x_4\ll 1$, the $y_0^2$ corrections will
additionally be suppressed by $O(x_4^2)$, and so the effective theory
becomes increasingly accurate in the Kaluza-Klein limit near to the
collision.  Close to $x_4=1$, the small-$y_0$ expansion fails hence
our results for the bulk geometry is no longer reliable.
For times $x_4>1$, the negative-tension brane no longer exists and the above
expression is not defined.

We can also ask what the physical counterpart of $t_4$, conformal time in the 
four-dimensional effective theory, is:
from (\ref{ad_b}), we find
\[
t_4 = \frac{b_+-b_-}{2 \bar{A}_4} = \frac{x_4}{y_0}-
\frac{y_0}{30 
     {( 1 - x_4^2) }^3}\Big( x_4^3 ( 5 - 14 x_4^2 + 5 x_4^4 )  - 
       5 x_4{( -1 + x_4^2 ) }^2 \ln (1 - x_4^2)\Big) + O(y_0^3).
\]
In comparison, the physical conformal times on the positive- and
negative-tension branes, defined via $b\, \d t_\pm = n\,\d t =
(n/y_0)(1-x_4^2)\,e^{-x_4^2/2}\,\d x_4 $, are, to $O(y_0^3)$,
\bea
t_+ &=& \frac{x_4}{y_0} + \frac{y_0}{30 
     {( 1 - x_4^2 ) }^3}\Big( 10 - 30x_4^2- x_4^3( 5 - 14 x_4^2 + 5 x_4^4 )  + 
       5 x_4 {( 1 - x_4^2 ) }^2 \ln (1 - x_4^2) \Big) \qquad  \\
t_- &=& \frac{x_4}{y_0} - \frac{y_0}{30 
     {( 1 - x_4^2 ) }^3}\Big( 10 -30 x_4^2+x_4^3(5 - 14 x_4^2 + 5 x_4^5 )  - 
       5 x_4 {( 1 - x_4^2 ) }^2 \ln (1 - x_4^2) \Big), \qquad
\eea
where we have used (\ref{ad_n}) and (\ref{ad_b}).

Remarkably, to lowest order in $y_0$, the two brane conformal times are
in agreement not only with each other, but also with the four-dimensional effective theory
conformal time.  Hence, in the limit in which $y_0^2$
corrections are negligible, there exists a universal four-dimensional time.
In this limit, $t_4=x_4/y_0$ and the brane scale factors are simply
given by $b_\pm = 1\pm\bar{A}_4 t_4 = 1\pm x_4$.
The four-dimensional effective scale factor, $b_4$, is given by
\[
\label{b4def}
(b_4)^2=b_+^2-b_-^2=4 \bar{A}_4 t_4 = 4 x_4^2.
\]


In order to describe the full five-dimensional geometry, 
one must specify the distance between
the branes $d$ as well as the metrics induced upon them. 
The distance between the branes is of particular interest
in the cyclic scenario, where an inter-brane force depending
on the inter-brane distance 
$d$ is postulated. In the lowest approximation,
where the branes are static, the four-dimensional effective theory
predicts that 
\[
\label{lncoth}
d =L \ln \coth\left({|\phi|\over\sqrt{6}}\right)=
L\ln\left(\frac{b_+}{b_-}\right) = L \ln \left({1+\bar{A}_4 t_4\over
  1-\bar{A}_4 t_4} \right). 
\]
Substituting our scaling solution and evaluating to leading order in 
$y_0$, we find
\[
\label{4d_pred}
d = L \ln \left(\frac{1+x_4}{1-x_4}\right)+O(y_0^2).
\]
(Again, this quantity is ill-defined for $x_4>1$).

In the full five-dimensional setup, 
a number of different measures of the inter-brane distance are
conceivable, and the inter-brane force could depend upon each 
of these, according to the precise higher-dimensional physics. 
One option would be to take the metric distance along
the extra dimension
\[
d_m =L \int_{-y_0}^{y_0} \sqrt{g_{yy}} \d y = L \int_{-y_0}^{y_0} nt
\d y = L \int_{-1}^{1} n x_4 e^{-\frac{1}{2}x_4^2}\d \w.
\]
Using (\ref{ad_n}), we obtain
\[
\label{d_m}
d_m = L \int_{-1}^{1}\frac{x_4}{1-\w x_4}\d \w+ O(y_0^2) = L \ln \left(\frac{1+x_4}{1-x_4}\right)+O(y_0^2),
\]
in agreement with (\ref{4d_pred}).

An alternative measure of the inter-brane distance is provided by
considering affinely parameterized spacelike geodesics running
from one brane to the other 
at constant Birkhoff-frame time
and noncompact coordinates $x^i$. 
The background interbrane distance
is just the affine parameter distance along
the geodesic, and the fluctuation in distance
is obtained by integrating the metric fluctuations
along the geodesic, as 
discussed in Appendix E.
One finds that, to leading order in $y_0$ only, the geodesic trajectories lie
purely in the $y$-direction.  Hence the affine distance
$d_a$ is trivially equal to the metric distance $d_m$ at leading
order, since
\[
d_a = L \int \sqrt{g_{ab}\dot{x}^a\dot{x}^b}\,\d \lambda = L \int nt\dot{y} \d \lambda = d_m,
\]
where the dots denote differentiation with respect to the affine parameter $\lambda$.
Both measures of the inter-brane distance therefore
coincide and are moreover in agreement with the four-dimensional
effective theory prediction, but only at leading order in $y_0$.

\subsection{Perturbations}

Since the four-dimensional Newtonian potential $\Phi_4$ represents the
\textit{anticonformal} part of the perturbed four-dimensional effective metric 
(see (\ref{4dmet})), it is unaffected by the conformal
factors in (\ref{map4}) relating the four-dimensional effective metric to the induced
brane metrics.  Hence we can directly compare the anticonformal part
of the perturbations of the induced metric on the branes, as calculated in five dimensions,
with $2\Phi_4$ in the four-dimensional effective theory.  The induced metric on the
branes is given by
\bea
\d s^2 &=& b^2\, \left( -(1+2 \Phi_L)\, \d t_\pm^2 + (1-2 \Psi_L )\, \d
\vec{x}^2 \right) \\
 &=& b^2\, (1+\Phi_L-\Psi_L) \left( -(1+\Phi_L+\Psi_L)\, \d t_\pm^2 + (1-
(\Psi_L+\Phi_L) )\, \d \vec{x}^2 \right),
\eea
where the background brane conformal time, $t_\pm$, is related to the bulk time via $b\, \d t_\pm = n\,\d t$. 
The anticonformal part of the metric perturbation is thus simply
$\Phi_L+\Psi_L$.
It is this quantity, evaluated on the branes to leading order in $y_0$, that we expect to
correspond to $2\Phi_4$ in the four-dimensional effective theory.

Using our results (\ref{adperts})
and (\ref{phi40soln}) from expanding about the scaling solution, we have to $O(y_0^2)$, 
\[
\label{thiseqn}
\frac{1}{2}(\Phi_L+\Psi_L)_+= \frac{1}{2}(\Phi_L+\Psi_L)_-=\phi_{40}(x_4)
= \frac{1}{x_4}\left(A_0 J_1(\tk x_4)+B_0 Y_1(\tk x_4)\right),
\]
with $A_0$ and $B_0$ as given in (\ref{A0B0}).
On the other hand, the Newtonian potential of the four-dimensional effective theory
is given by (\ref{phi4eft}).  Since $t_4$, the conformal time in the
four-dimensional effective theory, is related to the physical dimensionless brane
conformal time $x_4$ by $t_4=x_4/y_0$ (to lowest order in $y_0$), and moreover
$\tk=k/y_0$, we have $\tk x_4 = k t_4$.  Hence the four-dimensional effective theory
prediction for the Newtonian potential is in exact agreement with the
scaling solution holding at leading order in $y_0$, upon identifying
$\tA_0$ with $A_0/y_0$ and $\tB_0$ with $B_0/y_0$. 
The behavior of the Newtonian potential is
illustrated in Figure 8. 
\begin{figure}
\label{phi4plots}
\begin{tabular}{cc}
\begin{minipage}{8cm}
\hspace{-0.8cm}
\includegraphics[width=8cm]{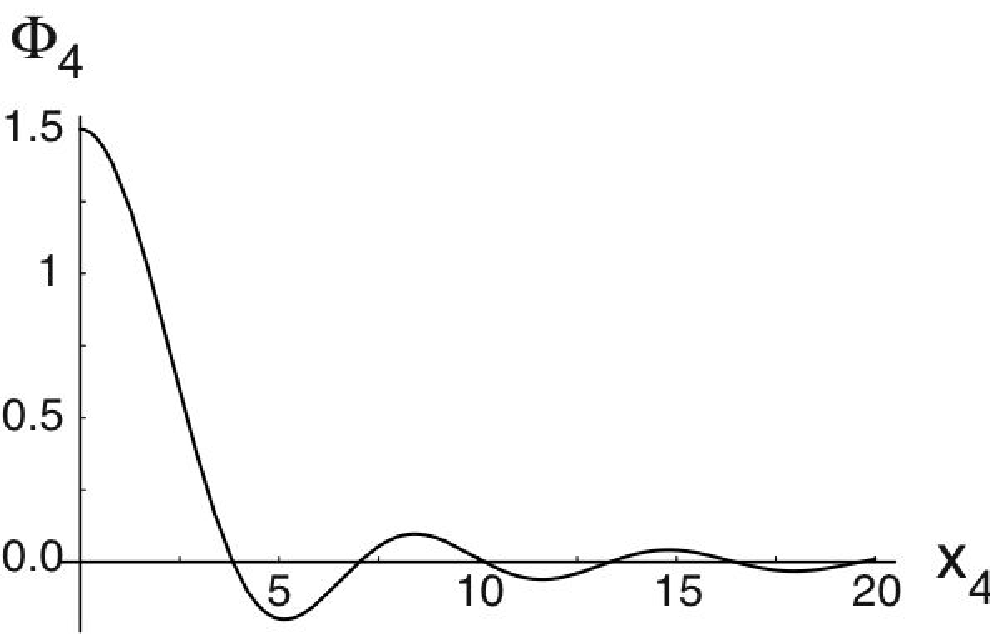}
\end{minipage}
\hspace{0cm}
\begin{minipage}{8cm}
\includegraphics[width=8cm]{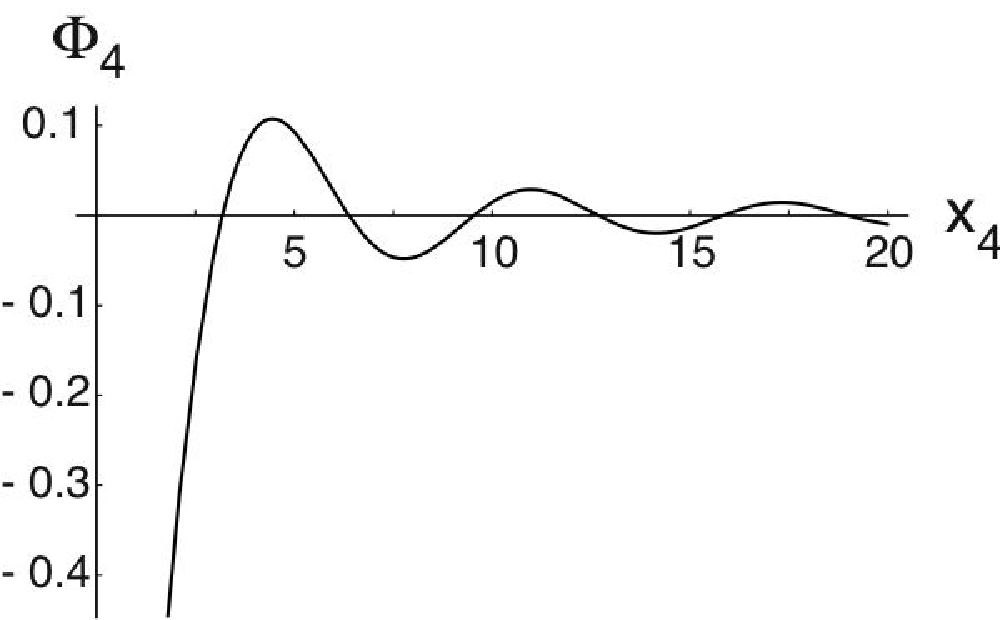}
\end{minipage}
\end{tabular}
\vspace{0.2cm}
\caption{
The four-dimensional Newtonian potential $\Phi_4$ on the
positive-tension brane, plotted to zeroth order in $y_0$ as a function
of the time $x_4$ for wavelength $\tk = 
1$.  The plot on the left illustrates the mode with $A=1$ and $\tB=0$,
while the plot on the right has $A=0$ and $\tB=1$.  
}
\end{figure}

Turning our attention now to the radion perturbation, $\delta \phi$, we know from our
earlier considerations that this quantity is related to the
perturbation $\delta d$ in the inter-brane separation.  Specifically,
from varying (\ref{lncoth}), we find
\[
\delta d = 2 L\, \cosech\left(\sqrt{\frac{2}{3}}\,\phi\right)\frac{\delta
  \phi}{\sqrt{6}}.
\]   
Inserting the four-dimensional effective theory predictions for $\phi$ and $\delta
\phi$, we obtain
\[
\label{deltad}
\frac{\delta d}{L}=\left(\frac{4 \bar{A}_4 t_4}{(\bar{A}_4 t_4)^2
  -1}\right)\left(\frac{2}{3}\,k\left(\tA_0
J_0(kt_4)+\tB_0Y_0(kt_4)\right) -\frac{1}{t_4}\left(\tA_0 J_1(kt_4)+\tB_0
Y_1(kt_4)\right) \right), 
\]
where to lowest order $\bar{A}_4 = y_0+O(y_0^3)$.

In comparison, the perturbation in the metric distance
between the branes is
\[
\label{deltad_m}
\frac{\delta d_m}{L} = \int_{-y_0}^{y_0}nt\Gamma_L \d y =
\int_{-1}^{1} \frac{n x_4 \xi_4}{b^2}\, e^{-\frac{1}{2}x_4^2} \d \w,
\]
where we have used (\ref{phi4xi4}).  Evaluating the integral using
(\ref{xi4soln}), to an accuracy of $O(y_0^2)$ we obtain
\bea
\frac{\delta d_m}{L}&=&\int_{-1}^{1} \frac{n x_4 \xi_{40}(x_4)}{b^2}\,
e^{-\frac{1}{2}x_4^2} \d \w
= \frac{2 x_4 \xi_{40}(x_4)}{(1-x_4^2)^2}
\nonumber \\ &=& 
\frac{1}{(x_4^2-1)}\left(\frac{8}{3}\,\tk x_4 (A_0 J_0(\tk x_4)+B_0
Y_0(\tk x_4))-4 (A_0 J_1(\tk x_4)+B_0 Y_1(\tk x_4))\right),
\eea
which is in agreement with (\ref{deltad}) when we set $\bar{A}_4 t_4
\sim y_0 t_4 = x_4$, along with $\tA_0=A_0/y_0$, $\tB_0=B_0/y_0$ and $k=\tk y_0$. 
The calculations in Appendix E show moreover that the perturbation
in the affine distance between the branes, $\delta d_a$, is 
identical to the perturbation in the metric distance $\delta d_m$,
to lowest order in $y_0$.  

The four-dimensional effective theory thus correctly predicts the Newtonian
potential $\Phi_4$ and the radion perturbation $\delta\phi$, but only
in the limit in which the $y_0^2$ corrections are
negligible and the bulk geometry is described by the scaling solution.  
While these corrections are automatically small at very
early or very late times, at intermediate times they cannot be
ignored and introduce effects that cannot be described by four-dimensional
effective theory.
The only five-dimensional longitudinal gauge metric perturbation we
have not used in any of the above is $W_L$: 
this component is effectively invisible to the four-dimensional effective
theory, since it vanishes on both branes and has no effect on the
inter-brane separation.  

\section{Mixing of Growing and Decaying Modes}

Regardless of the rapidity of the brane collision
$y_0$, one expects a four-dimensional effective description to hold
both near to the collision, when the brane separation is much less than an AdS
length, and also when the branes are widely separated over many AdS lengths.  In the
former case, the warping of the bulk geometry is negligible and a
Kaluza-Klein type reduction is feasible, and in the latter case, one
expects to recover brane-localized gravity.  At the transition between
these two regions, when the brane separation is of order one AdS
length, one might anticipate a breakdown of the four-dimensional effective description.

However, when the brane separation is of order a few AdS
lengths, the negative-tension brane reaches the horizon of the bulk
black hole and the small-$y_0$ expansion fails.  This failure hampers any efforts to
probe the breakdown of the four-dimensional effective theory at $x_4=1$ directly;
instead, we will look for evidence of mixing between the four-dimensional perturbation modes 
in the transition from Kaluza-Klein to brane-localized gravity.

To see this in action we have to compare the behavior of the
perturbations at very small times with that at very late times: in
both of these limits a four-dimensional effective description should apply,
regardless of the collision rapidity $y_0$, in which the four-dimensional Newtonian
potential $\Phi_4$ satisfies
\[
\Phi_4 = \frac{1}{x_4}\left(A_0 J_1(\tk x_4)+B_0 Y_1(\tk x_4)\right).
\]
Expanding this out on long wavelengths $\tk\ll 1$, taking in addition $\tk x_4 \ll 1$, we find
\[
\label{phi4exp}
\Phi_4 = -\,\frac{\tB_4^0}{x_4^2}+A_4^0+\frac{1}{2}\,\tB_4^0 \tk^2 \ln{\tk
  x_4}-\frac{1}{8}\,A_4^0 \tk^2 x_4^2+O(k^4),
\]
where the dimensionless constants $A_4^0$ and $\tB_4^0$ are given in terms of the five-dimensional 
perturbation amplitudes $A$ and $\tB$ by
\[
\label{A4B4}
A_4^0 = \frac{3}{2}\,A-\frac{9}{16}\,\tB-\frac{1}{8}\tB\tk^2, \qquad \tB_4^0
= \frac{1}{2}\tB,
\]
where we have used (\ref{A0B0}) and (\ref{thiseqn}), recalling that
$\Phi_4 = \phi_{40}$ at leading order in $y_0$.

In comparison, using our results from expanding about the scaling solution,
we find the Newtonian potential on the positive-tension brane at small
times $x_4 \ll 1$ is given by
\bea
\label{phi4early}
\Phi_4 &=&
\left(-\,\frac{\tB}{2 x_4^2}+
\frac{3}{2}\,A-\frac{9}{16}\,\tB-\frac{1}{8}\,
\tB \tk^2+\frac{1}{4}\, \tB \tk^2 \ln{\tk x_4} -\frac{3}{16}\,A\tk^2 x_4^2+\frac{9}{128}\,\tB \tk^2
x_4^2\right) \nonumber \\ &&
+y_0^2 \Bigg(\frac{11}{120}\,\tB-3 A x_4-\frac{47}{8}\,\tB
x_4-\frac{1}{2}\,\tB\tk^2 x_4 \ln{\tk x_4}+6Ax_4^2\nonumber \\ &&
\qquad\qquad +\frac{1084}{105}\,\tB x_4^2 -\frac{211}{960}\,\tB
\tk^2x_4^2+\tB\tk^2 x_4^2 \ln{\tk
  x_4}\Bigg) +O(x_4^3)+O(y_0^4).
\eea
Examining this, we see that to zeroth order in $y_0$ the result is in
exact agreement with (\ref{phi4exp}) and (\ref{A4B4}).  At $y_0^2$
order, however, extra terms appear that are not present in
(\ref{phi4exp}).  Nonetheless, at sufficiently small times the
effective theory is still valid as these `extra' terms are subleading
in $x_4$: in this limit we find
\[
\Phi_4 =
-\,\frac{\tB^E_4}{x_4^2}+A_4^E+\frac{1}{2}\,\tB_4^E\,\tk^2\ln{\tk
    x_4}+O(x_4) +O(\tk^4)
\]
(the superscript $E$ indicating early times), in accordance with the four-dimensional effective theory, where 
\[
A_4^E = A_4^0 + \frac{11}{120}\tB y_0^2, \qquad
\tB_4^E =\tB_4^0.
\]

At late times such that $x_4\gg 1$ (but still on sufficiently long
wavelengths that $\tk x_4 \ll 1$),
we find on the positive-tension brane
\bea
\label{phi4late}
\Phi_4 &=&
\left(-\,\frac{\tB}{2 x_4^2}+
\frac{3}{2}\,A-\frac{9}{16}\,\tB-\frac{1}{8}\,
\tB \tk^2+\frac{1}{4}\, \tB \tk^2 \ln{\tk x_4} -\frac{3}{16}\,A\tk^2 x_4^2+\frac{9}{128}\,\tB \tk^2
x_4^2\right) \nonumber \\ &&
+y_0^2 \Bigg(-\,\frac{A}{3 x_4^2}-\frac{\tB}{24
  x_4^2}-\frac{A\tk^2}{8x_4^2} +\frac{173\tB\tk^2}{960
  x_4^2}-\frac{\tB \tk^2 \ln{\tk}}{18x_4^2}-\frac{\tB \tk^2 \ln{
  x_4}}{12 x_4^2}\nonumber \\ &&\qquad
-\frac{3}{8}\,\tB+\frac{2}{9}\tB\tk^2+\frac{1}{6}\tB\tk^2\ln{x_4}
+\frac{3}{64}\tB\tk^2 x_4^2\Bigg)+O\left(\frac{1}{x_4^3}\right)+O(y_0^4).
\eea
To zeroth order in $y_0$, the results again coincide with the
effective theory prediction (\ref{phi4exp}) and (\ref{A4B4}).
However, at $y_0^2$ order extra terms not present in the
four-dimensional effective description once more appear.  In spite of
this, at sufficiently late times the effective description still holds as these `extra'
terms are suppressed by inverse powers of $x_4$ relative to the leading
terms, which are 
\[
\Phi_4 = A_4^L-\,\frac{\tB^L_4}{x_4^2} -\,\frac{1}{8}\,A_4^L\, \tk^2
x_4^2
+ O(\tk^2 \ln{\tk x_4})
\]
(where the superscript $L$ indicates late times),
in agreement with the four-dimensional effective theory.
(Since $x_4\gg 1$, we find $\tk \ll \tk x_4 \ll 1$, and we have chosen
to retain terms of $O(\tk^2 x_4^2)$ but to drop terms of $O(\tk^2)$.
The term of $O(x_4^{-2})$ is much larger than $O(\tk^2)$ and
so is similarly retained).
Fitting this to (\ref{phi4late}), we find \cite{comment3}
\[
A_4^L = A_4^0 -\frac{3}{8}\,\tB y_0^2, \qquad \tB_4^L =
\tB_4^0+\left(\frac{A}{3}+\frac{\tB}{24}\right)y_0^2.
\]

Comparing the amplitudes of the two four-dimensional modes
at early times, $A_4^E$ and $\tB_4^E$, 
with their counterparts $A_4^L$ and $\tB_4^L$ at late times, we see
clearly that the amplitudes differ at $y_0^2$ order.
Using (\ref{A4B4}), we find:
\[
\left(\begin{array}{c} A_4^L \\ \tB_4^L \end{array}\right) =
\left(\begin{array}{cc} 1 \  \ \ \ & -\frac{14}{15} y_0^2 \\
  \frac{2}{9}y_0^2 \ \ \ \ &
  1+(\frac{1}{3}+\frac{\tk^2}{18})y_0^2\end{array}\right)\left(\begin{array}{c} A_4^E \\ \tB_4^E \end{array}\right) .
\]
Hence 
the four-dimensional perturbation modes (as defined at very
early or very late times) undergo mixing.   

\section{Conclusion}

In this paper we have developed a set of powerful analytical methods
which, we believe, render braneworld cosmological perturbation
theory solvable. 

Considering the simplest possible cosmological scenario, consisting of slowly-moving,
flat, empty branes emerging from a collision, 
we have found a striking example of how the four-dimensional effective
theory breaks down at first nontrivial order in the brane speed.  
As the branes separate, a qualitative change in
the nature of the low energy modes occurs; from being nearly uniform
across the extra dimension when the brane separation is small, to
being exponentially localized on the positive-tension brane when the branes
are widely separated.  
If the branes separate at finite speed, the 
localization process fails to keep up with the brane separation and
the low energy modes do not evolve adiabatically.
Instead, a given Kaluza-Klein zero mode at early times will
generically evolve into a mixture of both brane-localized zero modes and excited modes
in the late-time theory.  
From the perspective of the four-dimensional theory, this is
manifested in the mixing of the four-dimensional effective
perturbation modes between early and late times, as we have calculated explicitly.
Such a mixing would be
impossible were a local four-dimensional effective theory to remain valid
throughout cosmic history: mode-mixing is literally a signature of
higher-dimensional physics, writ large across the sky.

As we show in a companion paper \cite{cyclicpaper}, this breakdown in the
four-dimensional effective description has further ramifications
for cosmology.  A key quantity of interest is the comoving curvature
perturbation $\zeta$ on the positive-tension brane, which is both gauge-invariant 
and, in the absence of additional bulk stresses, conserved on long wavelengths.
We show that, at first nontrivial order in the brane speed, $\zeta$
differs from its four-dimensional effective theory analogue, $\zeta_4$.
Hence, while the five-dimensional $\zeta$ is exactly conserved on long
wavelengths in the absence of bulk stresses, the four-dimensional
effective theory $\zeta_4$ is not precisely conserved, contrary to the
predictions of the four-dimensional effective theory. 
This has important implications for the propagation of perturbations before and after the bounce in 
cosmologies undergoing a big crunch/big bang transition, such as the ekpyrotic and cyclic models.

The methods developed in this paper should moreover readily extend to
other braneworld models; for example those
containing matter on the branes, and to models deriving from heterotic
M theory which are better motivated from a fundamental perspective \cite{Lukas}.
A further application would be to probe the dynamical
evolution of braneworld black holes \cite{EFT_BH} and black strings \cite{BS} in an expanding
cosmological background.

In conclusion, the strength of the expansion about the scaling
solution developed in
this paper lies in its ability to interpolate between very early and very
late time behaviors, spanning the gap in which the effective theory fails.
Not only can we solve for the full five-dimensional
background and perturbations of a colliding braneworld, 
but our solution takes 
us beyond the four-dimensional
effective theory and into the domain of intrinsically
higher-dimensional physics. 

\vspace{0.5cm}

\begin{center}
 ***
\end{center}

\textit{Acknowledgments:} We thank many colleagues for
discussions and useful comments, especially C.~de Rham, 
G.~Gibbons, J.-L.~Lehners, G.~Niz, A.~Tolley, S.~Webster and T.~Wiseman. 
PLM thanks PPARC and St.~John's College, Cambridge, for support. 
NT thanks PPARC and the Darley Fellowship at Downing College, Cambridge.
This work was also supported in part 
by US Department of Energy grant DE-FG02-91ER40671 (PJS).

\begin{appendix}

\section{Five-dimensional longitudinal gauge}

\subsection{Gauge invariant variables}

Starting with the background metric in the form (\ref{metrica}), 
the most general scalar metric perturbation can be written as \cite{Carsten}
\bea
\d s^2 &=& n^2\left(-(1+2\Phi)\,\d t^2 -2W\,\d t\, \d y + t^2 (1-2\Gamma)\,\d
  y^2 \right) - 2\nabla_i \alpha \,\d x^i \,\d t+2t^2\,\nabla_i\beta \,\d y \,\d
  x^i \nonumber \\
&& + b^2\left( (1-2\Psi)\,\delta_{ij}-2\nabla_i\nabla_j\chi\right)\d
  x^i\,\d x^j .
\eea
Under a gauge transformation $x^A\tt x^A+\xi^A$, these variables transform as
\bea
	\nonumber & &
	\Phi \rightarrow \Phi-\dot{\xi}^t-\xi^t \frac{\dot{n}}{n}-\xi^y \frac{n'}{n}, \\
	\nonumber & &
	\Gamma \rightarrow \Gamma+\xi'^y+\frac{1}{t}\xi^t+\xi^t 
	\frac{\dot{n}}{n}+\xi^y \frac{n'}{n}, \\
	\nonumber & &
	W \rightarrow W-\xi'^t+t^2 \dot{\xi}^y,  \\
	\nonumber & &
	\alpha \rightarrow \alpha-\xi^t+\frac{b^2}{n^2} \dot{\xi}^s, \\
	\nonumber & &
	\beta \rightarrow \beta-\xi^y-\frac{b^2}{n^2 t^2} \xi'^s,  \\
	\nonumber & &
	\Psi \rightarrow \Psi+\xi^t \frac{\dot{b}}{b}+\xi^y \frac{b'}{b}, \\
	& &
	\chi \rightarrow \chi+\xi^s , 
\eea
where dots and primes indicate differentiation with respect to $t$ and $y$ respectively. 
Since a five-vector $\xi^A$ has three scalar degrees of freedom $\xi^t$,
$\xi^y$ and $\xi^i=\nabla_i \xi^{s}$,  only four of the seven functions
$(\Phi,\Gamma,W,\alpha,\beta,\Psi,\chi)$ are physical. 
We can therefore construct four gauge-invariant variables, which are
\bea
\nonumber & &
\Phi_{\inv}=\Phi-\dot{\tilde{\alpha}}-\tilde{\alpha}\, \frac{\dot{n}}{n}-\tilde{\beta}\, \frac{n'}{n}, \\
\nonumber & &
\Gamma_{\inv}=\Gamma+\tilde{\beta}'+\frac{1}{t}\,\tilde{\alpha}
+\tilde{\alpha}\, \frac{\dot{n}}{n}+\tilde{\beta}\, \frac{n'}{n}, \\
\nonumber & &
W_{\inv}=W-\tilde{\alpha}'+t^2\, \dot{\tilde{\beta}}, \\
 & &
\Psi_{\inv}=\Psi+\frac{\dot{b}}{b}\,\tilde{\alpha}+\frac{b'}{b}\,
\tilde{\beta},
\label{gauges}
\eea
where $\tilde{\alpha}=\alpha-\frac{b^2}{n^2} \dot{\chi}$ and
$\tilde{\beta}=\beta+\frac{b^2}{n^2 t^2} \chi'$. 

In analogy with the four-dimensional case, we then define
five-dimensional longitudinal gauge by $\chi=\alpha=\beta=0$, giving
\begin{eqnarray}
	\nonumber & &
	\Phi_{\inv}=\Phi_L, \\
	\nonumber & &
	\Gamma_{\inv}=\Gamma_L, \\
	\nonumber & &
	W_{\inv}=W_L, \\
  	 & &
	\Psi_{\inv}=\Psi_L ,
\end{eqnarray}
\ie the gauge-invariant variables are equal to the values of the metric perturbations
in longitudinal gauge.
This gauge is spatially
isotropic in the $x^i$ coordinates, although in general there will be a
non-zero $t$--$y$ component of the metric.


\subsection{Position of branes}

In general, the locations of the branes will be different for
different choices of gauge.  In the case where the brane matter has no
anisotropic stresses this is easy to establish.  Working out the
Israel matching conditions, we find that $\beta$ on the branes is
related to the anisotropic part of the brane's stress-energy.
If we consider only perfect fluids, for which the shear vanishes, then
the Israel matching conditions give $\beta(y=\pm y_0)=0$.  

From the gauge transformations above, we can transform into the gauge $\alpha=\chi=0$
using only a $\xi^s$ and a $\xi^t$ transformation.  We may then pass to
longitudinal gauge ($\alpha=\beta=\chi=0$) with the transformation
$\xi^y=\tilde{\beta}$ alone.  Since $\beta$ (and hence $\tilde{\beta}$)
vanishes on the branes, $\xi^y$ must also vanish leaving the brane
trajectories unperturbed.  Hence, in longitudinal gauge the brane 
locations remain at their unperturbed values $y=\pm y_0$.
Transforming to a completely arbitrary gauge, we see that in general
the brane locations are given by 
\[
y = \pm y_0 - \tilde{\beta}.
\]

\section{Polynomial expansion: Background}

Using the expansion in Dirichlet/Neumann
polynomials presented in Section IV to solve for the
background geometry, we find
\bea
N_0 &=& {1\over t}-{1\over 2}\,t\,y_0^2+{1\over 24}\,t\,(8-9\,t^2)\,y_0^4-{1\over
  720}\,t\,(136+900\,t^2+375\,t^4)\,y_0^6 \nonumber \\
&& +{1\over 40320}\,t\,(3968+354816\,t^2-348544\,t^4-36015\,t^6)\,y_0^8+O(y_0^{10})\\
N_3 &=& -\frac{1}{6} + \left(\frac{5}{72} - 2\,t^2 \right)\,y_0^2 - 
  {1\over 2160}\left( 61 - 20880\,t^2 + 19440\,t^4 \right)\,y_0^4 \nonumber \\
&& + \left( \frac{277}{24192} - \frac{743\,t^2}{20} + \frac{677\,t^4}{6}
  - \frac{101\,t^6}{3} \right)\,y_0^6+O(y_0^{8}) \\
N_4 &=& \frac{3\,t^3\,y_0^2}{2} + \frac{t^3\,\left( -28 + 33\,t^2 \right) \,y_0^4}{4} + 
  \frac{t^3\,\left( 1984 - 6776\,t^2 + 2715\,t^4 \right)
    \,y_0^6}{80}+O(y_0^8) \\
N_5 &=& -\frac{1}{120} - \frac{\left( -7 + 1800\,t^2 + 540\,t^4
  \right) \,y_0^2}{1800} \nonumber \\
&& - \frac{\left( 323 - 990528\,t^2 + 2207520\,t^4 + 362880\,t^6 \right)
    \,y_0^4}{201600}+O(y_0^6) \\
N_6 &=& t^3\,y_0^2 + \frac{1}{30}t^3\,\left( -142 + 371\,t^2 \right)
\,y_0^4+O(y_0^6) \\
N_7 &=& -\frac{1}{5040}- \frac{\left( -9 + 20384\,t^2 + 23520\,t^4
  \right) \,y_0^2}{94080}+O(y_0^4) \\
N_8 &=& \frac{3\,t^3\,y_0^2}{10}+O(y_0^4) \\
N_9 &=& \frac{1}{362880} +O(y_0^2) \\
N_{10} &=& O(y_0^2) 
\eea
and
{\allowdisplaybreaks
\bea
q_0 &=& 1 - \frac{3\,t^2\,y_0^2}{2} + \left( t^2 -
\frac{7\,t^4}{8} \right) \,y_0^4 +  
  \left( \frac{-17\,t^2}{30} + \frac{17\,t^4}{12} - \frac{55\,t^6}{48}
  \right) \,y_0^6 \nonumber \\
&& + \left( \frac{31\,t^2}{105} - \frac{9\,t^4}{5} + \frac{233\,t^6}{90} -
  \frac{245\,t^8}{128} \right) \,y_0^8 +  O(y_0^{10}) 
\\
q_3 &=&-2\,t^3\,y_0^2 + \left( \frac{29\,t^3}{3} - 8\,t^5 \right) \,y_0^4 + 
  \left( -\frac{743\,t^3}{20} + \frac{322\,t^5}{3} - 27\,t^7 \right)
  \,y_0^6+O(y_0^8)
\\ 
q_4 &=& \frac{t^4\,y_0^2}{2} + \left( \frac{-5\,t^4}{3} +
\frac{9\,t^6}{4} \right) \,y_0^4+O(y_0^6)
\\
q_5&=& - t^3\,y_0^2  + \left( \frac{737\,t^3}{150} - \frac{58\,t^5}{5}
\right) \,y_0^4+O(y_0^6)
\\
q_6 &=& \frac{t^4\,y_0^2}{3}+O(y_0^4)
\\
q_7 &=& \frac{-13\,t^3\,y_0^2}{60}+O(y_0^4)   
\\
q_8 &=& O(y_0^2)
\\
q_9 &=& O(y_0^2),
\eea
}
where we have set $L=1$ for clarity.  (To restore $L$ simply replace
$t \tt t/L$). 
The solution has been checked to satisfy all the remaining
Einstein equations explicitly.

\section{Polynomial expansion: Perturbations}

\subsection{All wavelengths}

Throughout this Appendix, we shall
set the AdS radius $L$ to unity. It is then simple
to restore $L$ by dimensions where needed, \ie  by
setting $t \tt t/L$ and $k \tt kL$. Note that
the coordinate $y$ is dimensionless.

Using the Dirichlet/Neumann polynomial expansion 
to solve for the perturbations, the
solution may be expressed in terms of the original longitudinal gauge
variables as
\[
\Phi_L = \mathcal{P}_\Phi^{(0)}(y,t)F^{(0)}(t)+ \mathcal{P}_\Phi^{(1)}(y,t)F^{(1)}(t)
\]
where
\[
F^{(n)}(t) =
\bar{A} J_n(kt)+\bar{B} Y_n(kt)
\]
for $n=0,\,1$ and $\gamma = 0.577\dots$ is the Euler-Mascheroni
constant.  The constants $\bar{A}$ and $\bar{B}$ are arbitrary functions of
$k$.  In order to be consistent with the series expansion in $t$
presented in Section III, we must set
\bea
\bar{A} &=& 12\,A+2\,B\,k^2\,(\ln{2}-\gamma ) -
\frac{9\,B\,y_0^2}{2} + \frac{233\,B\,y_0^4}{45}+O(y_0^6)
 \\
\bar{B} &=& B\, k^2\,\pi+O(y_0^6).
\eea

The polynomials $\mathcal{P}_\Phi^{(n)}$ are then given (for all $k$
and $t$) by
\bea
\mathcal{P}_\Phi^{(0)}(t,y) &=& -\frac{1}{6} + \frac{t\, 
      y}{3} + \frac{t^2\, \left(-2\, y^2 + y_0^2 \right)}{12} +
  \frac{t\, 
      y\, \left(11\, y^2 + 3\, \left(-11 + 3\, t^2 \right)\, 
          y_0^2 \right)}{36} + \nonumber \\
&&   \frac{t^2\, \left(-525\, y^4 + 90\, \left(19 - 5\, t^2 \right)\, y^2\, 
          y_0^2 +
            \left(511 - 180\, t^2 + 45\, k^2\, t^4 \right)\, 
          y_0^4 \right)}{2160} - \nonumber \\
&&  \frac{t\,y}{2160}\,\Bigg(3\,\big(-92 + (9 + 4\, k^2)\, t^2 \big)\, 
          y^4 + 30\, \big(92 - (219 + 4\, k^2)\, t^2 + k^2\, 
                t^4 \big)\, y^2\, y_0^2 + \nonumber \\ 
&&            \big(-6900 + (20087 + 300\, k^2)\, 
                  t^2 - 90\, (13 + k^2)\, t^4 + 90\, k^2\, 
                  t^6 \big)\, y_0^4\Bigg) +O(y_0^6),\nonumber \\
\eea
\bea
\mathcal{P}_\Phi^{(1)}(t,y) &=& \frac{1}{k\,t}\Bigg[\frac{1}{2} -
  \frac{t\,y}{2} + \frac{t^2\, \left(3\, y^2 + \left(-3 + k^2\, t^2 \right)\, 
          y_0^2 \right)}{12} - \nonumber \\
&&  \frac{t\,y\, \left( \left(3 + k^2\, t^2 \right)\, 
          y^2 + 3\, \left(-3 + \left(3 - k^2 \right)\, t^2 + 2\, k^2\, 
                t^4 \right)\, y_0^2 \right)}{36} + \nonumber \\
&&\frac{t^2}{2160}\Bigg(75\, (6 + k^2\, 
                t^2 )\, 
          y^4 + 90\, \left(-12 + 3\, (2 - k^2)\, t^2 + 2\, k^2\, 
                t^4 \right)\, y^2\, y_0^2 + \nonumber \\
&&            \left(718 + k^2\, t^2\, (-101 + 225\, t^2) \right)\, 
          y_0^4 \Bigg)- \nonumber \\
&& \frac{t\,y}{2160}\,\Bigg(3\,\left(3 + 2\, (-9 + 31\, k^2)\, t^2 + 2\, 
                k^2\, t^4 \right)\, 
          y^4 +\nonumber \\
&& 30\,\left(-3 + (219 - 62\, k^2)\, t^2 + 16\, 
                k^2\, t^4 \right)\, y^2\, y_0^2 + \nonumber \\
&&            \left(225 + (-20104 + 4650\, k^2)\, 
                  t^2 + (1215 - 1822\, k^2)\, t^4 + 765\, k^2\, 
                  t^6 \right)\, 
          y_0^4 \Bigg) \Bigg]\nonumber \\ && +O(y_0^6) . 
\eea
Since the $F^{(n)}$ are of zeroth order in $y_0$, the solution for
$\Phi_L$ to a given order less than $O(y_0^6)$ is found simply by truncating the
polynomials above.  (Should they be needed, results up to
$O(y_0^{14})$ can in addition be found at \cite{Website}). 

In a similar fashion we may express the solution for $\Psi_L$ as
\[
\Psi_L = \mathcal{P}_\Psi^{(0)}(y,t)F^{(0)}(t)+ \mathcal{P}_\Psi^{(1)}(y,t)F^{(1)}(t),
\]
where $F^{(n)}$ is defined as above and
\bea
\mathcal{P}_\Psi^{(0)}(t,y) &=& 
\frac{1}{6} - \frac{t\, y}{3} + \frac{t^2\, \left(y^2 + y_0^2 \right)}{6} +
  \frac{t\, 
      y\, \left(-2\, y^2 + 3\, \left(2 - 3\, t^2 \right)\, 
          y_0^2 \right)}{36} +\nonumber \\
&&  \frac{t^2\, \left(120\, y^4 + 450\, \left(-2 + t^2 \right)\, y^2\, y_0^2 +
            \left(644 + 450\, t^2 - 45\, k^2\, t^4 \right)\, 
          y_0^4 \right)}{2160} + \nonumber \\
&&  \frac{t\, 
      y}{2160} \Bigg(3\, \left(-2 + (9 + 4\, k^2)\, t^2 \right)\, 
          y^4 + 30\, \left(2 + 4\, (9 - k^2)\, t^2 + k^2\, 
                t^4 \right)\, y^2\, y_0^2 + \nonumber \\
&&            \left(-150 + (-2863 + 300\, k^2 )\, 
                  t^2 - 90\, (13 + k^2)\, t^4 + 90\, k^2\, 
                  t^6 \right)\, y_0^4 \Bigg)+O(y_0^6), \nonumber \\
\eea
\bea
\mathcal{P}_\Psi^{(1)}(t,y) &=& \frac{1}{k\,t}\Bigg[
\frac{t\,y}{2} - \frac{t^2\, \left(3\, y^2 + \left(3 + k^2\, t^2 \right)\, 
          y_0^2 \right)}{12} + \nonumber \\
&&  \frac{t\,y\, \left( \left(3 + k^2\, t^2 \right)\, 
          y^2 + 3\, \left(-3 + \left(3 - k^2 \right)\, t^2 + 2\, k^2\, 
                t^4 \right)\, 
          y_0^2 \right)}{36} - \nonumber \\
&& \frac{t^2}{2160}\Bigg(15\, (12 + 5\, k^2\, 
                t^2 )\, 
          y^4 + 90\, \left(-6 - 3\, (-2 + k^2)\, t^2 + 2\, k^2\, 
                t^4 \right)\, y^2\, y_0^2 +\nonumber \\
&&            \left(-752 + (540 - 101\, k^2)\, t^2 + 360\, k^2\, 
                  t^4 \right)\, y_0^4 \Bigg) + \nonumber \\
&&  \frac{t\,y}{2160}\Bigg( \left(9 + 6\, \left(-9 + k^2 \right)\, t^2 + 6\, k^2\, 
                t^4 \right)\, 
          y^4 + \nonumber \\
&& 30\, \left(-3 - \left(33 + 2\, k^2 \right)\, t^2 + 7\, k^2\, 
                t^4 \right)\, y^2\, y_0^2 + \nonumber \\
&&            \left(225 + 2\, \left(1288 + 75\, k^2 \right)\, 
                  t^2 + \left(1215 - 1012\, k^2 \right)\, t^4 + 765\, k^2\, 
                  t^6 \right)\, y_0^4 \Bigg)
\Bigg]\nonumber \\ && +O(y_0^6) . 
\eea

Finally, writing
\[
W_L = \mathcal{P}_W^{(0)}(y,t)F^{(0)}(t)+ \mathcal{P}_W^{(1)}(y,t)F^{(1)}(t),
\]
we find
\bea
\mathcal{P}_W^{(0)}(t,y) &=&-\frac{1}{60}\,t^2\, (y^2 - y_0^2)\,
\Big(-30 + 30\, t\, 
    y - 25\, \left(y^2 + (-5 + 3\, t^2)\, y_0^2 \right)
    +\nonumber \\
&& t\,  y\, \left(21\, y^2 +(-149 + 75\, t^2)\, y_0^2 \right)
    \Big)+O(y_0^6), \\
\nonumber \\
\mathcal{P}_W^{(1)}(t,y) &=&-\frac{1}{60\,k}\,t^2\, (y^2 - y_0^2)\,\Big(30\, y - 5\, 
k^2\, t\, \left(2\, y^2 + (-10 + 3\, t^2)\, y_0^2 \right)
+ \nonumber \\
&&  y\,\left( (12 + 11\, k^2\, t^2)\, 
    y^2 + \left(-38 + (60 - 69\, k^2)\, t^2 + 15\, k^2\, 
            t^4 \right)\,y_0^2\right)\Big) +O(y_0^6) . \nonumber \\ 
\eea

\subsection{Long wavelengths}

On long wavelengths, $F^{(n)}$ reduces to
\bea
F^{(0)}(t) &=& 12\,A - \frac{9\,B\,y_0^2}{2} + \frac{233\,B\,y_0^4}{45} +O(k^2)+O(y_0^6), \\
F^{(1)}(t) &=& \left( 6\,A\,t - \left( \frac{2}{t} +
\frac{9\,t\,y_0^2}{4} - \frac{233\,t\,y_0^4}{90}\right)\,B  \right)\,k
+O(k^2)+O(y_0^6).
\eea
For convenience, we provide here a separate listing of the metric
perturbations truncated at $O(k^2)$: 
\bea
\Phi_L &=& \big(A - \frac{B}{t^2}\big) + \big( \frac{B y}{t} + A t y \big)    + 
  \frac{1}{8}\left( B ( y_0^2 - 4 y^2 )  - 4 A t^2 ( y_0^2 + y^2 )
  \right) \nonumber \\ && + 
  \frac{y}{24 t}\big( B ( 3 y_0^2 ( -4 + t^2 )  + 4 y^2 )  + 
       4 A t^2 ( 3 y_0^2 ( -19 + 3 t^2 )  + 19 y^2 \big)  \nonumber \\
       && + 
  \Big( \frac{1}{6} A t^2 ( y_0^4 ( 29 - 6 t^2 )  +
    3 y_0^2 ( 13 - 2 t^2 )  y^2 - 10 y^4 )\nonumber \\ &&  - 
     \frac{1}{240}(B ( y_0^4 ( 56 - 45 t^2 )  +
       15 y_0^2 ( -16 + 5 t^2 )  y^2 + 100 y^4 )
     ) \Big) \nonumber \\ &&  
     + \frac{y}{240 t} \big( 2 A t^2 ( 5 y_0^4 ( 905 - 1338 t^2 + 75 t^4 )  + 
          10 y_0^2 ( -181 + 219 t^2 )  y^2 + 181 y^4 )  \nonumber \\
	  && + 
       B ( y_0^4 ( 50 - 3509 t^2 + 135 t^4 )  +
       5 y_0^2 ( -4 + 235 t^2 )  y^2 +  
          ( 2 - 12 t^2 )  y^4 )\big) \nonumber \\ && +  O(y_0^6), \\ 
\Psi_L &=& 2 A - \frac{y}{t}( B + A t^2 ) +
\frac{1}{4}\big( 2 A t^2 ( y_0^2 + y^2 )  +  
       B ( -y_0^2 + 2 y^2 )  \big)  \nonumber \\ && - 
  \frac{y}{24 t} \big( 4 A t^2 ( 3 y_0^2 ( -1 + 3 t^2 )  + y^2 )  + 
       B ( 3 y_0^2 ( -4 + t^2 )  + 4 y^2 )  \big) \nonumber \\ && + 
  \frac{1}{48}\Big( 8 A t^2 ( 2 y_0^4 ( 17 + 3 t^2 )  +
    3 y_0^2 ( -7 + 2 t^2 )  y^2 + y^4 )  \nonumber \\ && +  
       B ( y_0^4 ( 8 + 15 t^2 )  + 3 y_0^2 ( -8
       + 5 t^2 )  y^2 + 8 y^4 )  \Big)\nonumber \\ &&  -  
  \frac{y}{240 t} \Big( 2 A t^2 ( 25 y_0^4 ( 1 + 42 t^2 +
    15 t^4 )  - 10 y_0^2 ( 1 + 39 t^2 )  y^2 + y^4 
          )  \nonumber \\ && + B ( y_0^4 ( 50 + 721 t^2 + 135 t^4
	  )  - 5 y_0^2 ( 4 + 47 t^2 )  y^2 +  
          ( 2 - 12 t^2 )  y^4 )  \Big)  \nonumber \\ && +  O(y_0^6), \\ 
W_L &=& 
6 A t^2 ( -y_0^2 + y^2 )  - t ( B + 3 A t^2
)  y ( -y_0^2 + y^2 )  \nonumber \\ &&  +  
  \frac{1}{4}t^2 ( -y_0^2 + y^2 )  ( -9 B y_0^2 +
    20 A ( y_0^2 ( -5 + 3 t^2 )  + y^2 )
    )\nonumber \\ &&  - \frac{1}{120} t y ( -y_0^2 + y^2 )  (
     120 A t^2 ( y_0^2 ( -26 + 9 t^2 )  + 3 y^2
     ) \nonumber \\ &&  +  
       B ( y_0^2 ( -152 + 105 t^2 )  + 48 y^2 )  ) +  O(y_0^6).
\eea
Terms up to $O(y_0^{14})$ are available at \cite{Website}.

\section{Perturbations from expansion about the scaling \indent\indent\indent\indent\indent\indent solution}

Following the method presented in Section V, the
perturbations were computed to $O(y_0^4$). On long wavelengths,
the five-dimensional longitudinal gauge variables 
take the form
\bea
\Phi_L &=&  f^\Phi_0+y_0^2 (f^\Phi_1+f^\Phi_2
\ln{(1+x_4)} + f^\Phi_3 \ln{(1-x_4)}+f^\phi_4 \ln{(1-\w x_4)}, \\
\Psi_L &=&  f^\Psi_0+y_0^2 (f^\Psi_1+f^\Psi_2
\ln{(1+x_4)} + f^\Psi_3 \ln{(1-x_4)}+f^\Psi_4 \ln{(1-\w x_4)}, \\
W_L&=& e^{-\frac{1}{2}x_4^2}\left( f^W_0+y_0^2 (f^W_1+f^W_2
\ln{(1+x_4)} + f^W_3 \ln{(1-x_4)}+f^W_4 \ln{(1-\w x_4)}\right),
\eea
where the $f$ are rational functions of $x_4$ and $\w$.  For $\Phi_L$,
we have
\bea
f^\Phi_0 &=& \frac{1}{16 x_4^2 ( -1 + x_4^2 ) }\Big(16 \tB - 16 \tB \w x_4 - 2 ( 8 A + \tB -
  4 \tB \w^2 )  x_4^2 + 2 ( -8 A + 3 \tB )  \w x_4^3 \nonumber \\ && +  
    ( 8 A - 3 \tB )  ( 3 + \w^2 )
     x_4^4\Big),\\
f^\Phi_1 &=& \frac{1}{960 x_4^4 {( -1 + x_4^2 ) }^5}\Big(8 A x_4^5 \big(
-580 x_4 + 95 x_4^3 - 576 \w^5 x_4^4 + 281 x_4^5 + 96 \w^6 x_4^5 - 60
x_4^7 \nonumber \\ && + 
       5 \w^4 x_4 ( 40 + 167 x_4^2 - 39 x_4^4 )  + 20 \w^3 ( -19 - 5
       x_4^2 + 48 x_4^4 ) \nonumber \\ && +  
       10 \w^2 x_4 ( -78 - 117 x_4^2 + x_4^4 + 2 x_4^6 )  +
       4 \w ( 285 + 100 x_4^2 - 25 x_4^4 - 29 x_4^6 + 5 x_4^8
       )  \big)  \nonumber \\ && +  
    \tB \big( -1920 + x_4 ( 480 \w + 80 ( 91 + 15 \w^2
    )  x_4 \nonumber \\ && - 160 \w ( 5 + 4 \w^2 )  x_4^2 +  
          40 ( -273 - 116 \w^2 + 7 \w^4 )  x_4^3 - 20 \w ( -649 + 127
	  \w^2 )  x_4^4 \nonumber \\ && + 
          4 ( 1231 - 3285 \w^2 + 2060 \w^4 )  x_4^5 -
	  4 \w ( 2036 - 2115 \w^2 + 1152 \w^4 )  x_4^6 \nonumber \\ &&
	  +  
          ( 1107 + 8102 \w^2 - 4905 \w^4 + 768 \w^6 )  x_4^7 + 
          12 \w ( 233 + 48 \w^2 ( -5 + 3 \w^2 )  )  x_4^8 \nonumber \\
	  && - 
          ( 3131 + 1582 \w^2 - 585 \w^4 + 288 \w^6 )  x_4^9
	  - 772 \w x_4^{10} + 20 ( 85 + 29 \w^2 )  x_4^{11}
	 \nonumber \\ && +  
          180 \w x_4^{12}  - 120 ( 3 + \w^2 )  x_4^{13}
	  )  \big) \Big), \\ 
f^\Phi_2 &=& \frac{1}{48 x_4^5 {( -1 + x_4^2 ) }^3}\Big(\tB\big( 48 +
( 36 - 48 \w )  x_4 + 12 ( -7 - 6 \w + 2 \w^2 )
 x_4^2 \nonumber \\ && +  
       4 ( -13 + 6 \w + 9 \w^2 )  x_4^3 + ( 60 + 80 \w
       - 12 \w^2 )  x_4^4 - 4 ( -8 - 6 \w + 9 \w^2 )
        x_4^5 \nonumber \\ && -  
       3 ( 8 + 7 \w + 4 \w^2 )  x_4^6 + ( -11 + 5 \w^2 )  x_4^7 + 3 \w x_4^8 \big)  + 
    8 A x_4^6 ( x_4 + \w^2 x_4 \nonumber \\ && - \w ( 1 + x_4^2 )  )
     \Big), 
\eea
\bea
f^\Phi_3 &=& \frac{1}{48 x_4^5 {( -1 + x_4^2 )
  }^3}\Big(\tB\big( -48 + 12( 3 + 4\w )  x_4 - 12 ( -7 + 6 \w + 2 \w^2
)  x_4^2 \nonumber \\ && + 
       4 ( -13 - 6 \w + 9 \w^2 )  x_4^3 + 4 ( -15 + 20 \w + 3 \w^2 )  x_4^4 - 
       4 ( -8 + 6 \w + 9 \w^2 )  x_4^5 \nonumber \\ && + 3 ( 8 - 7 \w
       + 4 \w^2 )  x_4^6 + ( -11 + 5 \w^2 )  x_4^7 +
       3 \w x_4^8 
       )  \nonumber \\ && + 8 A x_4^6 ( x_4 + \w^2 x_4 - \w ( 1 + x_4^2
       )  ) \Big), \\
f^\phi_4 &=& \frac{3 \tB {( -1 + \w x_4 )
  }^2}{2 x_4^4 {( -1 + x_4^2 ) }^2} .
\eea

Similarly, for $\Psi_L$, we find
{\allowdisplaybreaks
\bea
f^\Psi_0 &=& \frac{1}{16 x_4 ( -1 + x_4^2 ) }\Big(16 \tB \w - 4 ( 8 A + \tB ( -1 + 2 \w^2
  )  )  x_4 + 2 ( 8 A - 3 \tB )  \w x_4^2 \nonumber \\ && +  
    ( -8 A + 3 \tB )  ( -3 + \w^2 )
     x_4^3\Big), \\
f^\Psi_1 &=& \frac{1}{960 x_4^3 {( -1 +
    x_4^2 ) }^5}\Big(-480 \tB \w - 240 \tB ( -7 + 5 \w^2 )  x_4
  + 160 \tB \w ( 5 + 4 \w^2 )  x_4^2 \nonumber \\ && -  
    40 \tB ( 143 - 104 \w^2 + \w^4 )  x_4^3 + 20 \w 
     ( \tB ( 197 - 155 \w^2 )  + 8 A ( -3 + \w^2 )  )  x_4^4 \nonumber
    \\ && + 
    20 ( -8 A ( 34 - 21 \w^2 + \w^4 )  + \tB (
    98 - 369 \w^2 + 101 \w^4 )  )  x_4^5\nonumber \\ && -  
    4 \w ( 200 A ( -14 + 5 \w^2 )  +  \tB ( 34 -
    975 \w^2 + 288 \w^4 )  )  x_4^6 +  
    ( 40 A ( 55 - 234 \w^2 + 67 \w^4 ) \nonumber \\ &&  + \tB (
    2455 + 838 \w^2 - 1005 \w^4 + 192 \w^6 )  )  x_4^7 -  
    4 \w ( 3 \tB ( 53 + 120 \w^2 - 36 \w^4 ) \nonumber \\ &&  +
    8 A ( 155 - 120 \w^2 + 36 \w^4 )  )  x_4^8 +  
    ( \tB ( -3515 + 1042 \w^2 + 225 \w^4 - 72 \w^6 )  \nonumber \\ && +
    8 A ( 89 + 170 \w^2 - 75 \w^4 + 24 \w^6 )  )
     x_4^9 +  
    4 ( 232 A + 193 \tB )  \w x_4^{10} - 20 (
    8 A ( 1 + \w^2 )  \nonumber \\ && + \tB ( -91 + 29 \w^2 )
    )   
     x_4^{11} - 20 ( 8 A + 9 \tB )  \w x_4^{12} +
     120 \tB ( -3 + \w^2 )  x_4^{13}\Big), \\
f^\Psi_2 &=& \frac{-1}{48 x_4^4  
    {( -1 + x_4^2 ) }^3}\Big( 8 A x_4^5 ( x_4 + \w^2 x_4 - \w ( 1 + x_4^2 )  )  + 
      \tB ( 36 + 60 x_4 - 36 x_4^2 - 84 x_4^3 +\nonumber \\ &&  24 x_4^5 + 5 x_4^6 +
      \w^2 x_4 ( 24 + 36 x_4 - 12 x_4^2 - 36 x_4^3 - 12 x_4^4 + 5 x_4^5
      ) \nonumber \\ && +  
         \w ( -48 - 72 x_4 +  24 x_4^2 + 80 x_4^3 + 24 x_4^4 - 21 x_4^5
      + 3 x_4^7 )  )   \Big),
\\
f^\Psi_3 &=& \frac{1}{48 x_4^4 {( -1 + x_4^2 ) }^3}\Big(8 A x_4^5 ( - x_4 - \w^2 x_4 + \w (1+x_4^2) )
   + 
    \tB ( -36 + 60 x_4 + 36 x_4^2 - \nonumber \\ && 84 x_4^3 +  24 x_4^5 - 5 x_4^6 -
      \w^2 x_4 ( -24 + 36 x_4 + 12 x_4^2 - 36 x_4^3 + 12 x_4^4 +
      5 x_4^5 ) \nonumber \\ && +  
       \w ( -48 +  72 x_4 + 24 x_4^2 - 80 x_4^3 + 24 x_4^4 + 21 x_4^5 -
      3 x_4^7 )  ) \Big),
 \\
f^\Psi_4 &=& \frac{-3 \tB {( -1 + \w x_4 )
  }^2}{2 x_4^4 {( -1 + x_4^2 ) }^2} .
\eea
}

Finally, for $W_L$, we have
\bea
f^W_0 &=&\frac{( -1 + \w^2 )  x_4 ( -24 A x_4 (
  -2 + \w x_4 )  + \tB ( -18 x_4 + \w ( -8 + 9 x_4^2
  )  ) 
      ) }{8 {( -1 + x_4^2 ) }^2},\\
f^W_1 &=&\frac{(-1 + \w^2)}{480 x_4^2 {( -1 + x_4^2 ) }^7}  \Big( 8 A x_4^4 \big( 1500
  + 84 \w^4 x_4^4 ( -12 + x_4^2 ) \nonumber \\
	 &&  -  
         6 \w^2 ( 50 + 100 x_4^2 - 427 x_4^4 + x_4^6 )  +
	 3 \w^3 x_4 ( 60 + 585 x_4^2 - 160 x_4^4 + 3 x_4^6 )
	  \nonumber \\ &&+  
         \w^5 ( 168 x_4^5 - 36 x_4^7 )  - 6 x_4^2 ( -590
	 + 243 x_4^2 + 201 x_4^4 - 4 x_4^6 + 10 x_4^8 )  \nonumber \\
	 && +  
         \w x_4 ( -1560 - 4935 x_4^2 + 2000 x_4^4 - 265 x_4^6 + 92
	 x_4^8 )  \big)  \nonumber \\ && + 
      \tB \big( 1440 + x_4 ( -84 \w^4 x_4^5 ( 24 + 28 x_4^2 +
      3 x_4^4 )  \nonumber \\ && + 12 \w^5 x_4^6 ( 40 + 6 x_4^2 + 9 x_4^4
      )  +  
            6 \w^2 x_4^3 ( -450 + 964 x_4^2 + 863 x_4^4 + 3 x_4^6 )
	    \nonumber \\ && - 
            3 \w^3 x_4^2 ( -40 - 748 x_4^2 - 2333 x_4^4 + 672 x_4^6 +
	    9 x_4^8 )  \nonumber \\ && - 
            6 x_4 ( 2160 - 5490 x_4^2 + 3770 x_4^4 - 2249 x_4^6 +
	    597 x_4^8 - 588 x_4^{10} + 90 x_4^{12} ) \nonumber \\ &&  +  
            \w ( -2160 + 18920 x_4^2 - 53216 x_4^4 + 20629 x_4^6 -
	    11216 x_4^8 + 5579 x_4^{10} \nonumber \\ && - 2236 x_4^{12} + 360 x_4^{14}
	    )  )  
         \big)  \Big), \\
f^W_2 &=& \frac{( 1 - \w )}{24 {( x_4 - x_4^3 ) }^4}  \Big( \tB \big( -144 + 108 ( -1 + 2 \w )  x_4 - 
           24 ( -5 + \w )  ( 3 + 2 \w )  x_4^2 \nonumber \\ && -
	   36 ( -11 + \w ( 16 + \w )  )  x_4^3 +  
           24 \w ( -29 + 7 \w )  x_4^4 \nonumber \\ && + 36 ( -2 +
	   \w ( -2 + 7 \w )  )  x_4^5 +  
           3 ( 1 + \w )  ( 3 + 32 \w )  x_4^6 +
	   7 \w ( 1 + \w )  x_4^7 \nonumber \\ &&+ 27 ( 1 + \w )
	    x_4^8  -  
           9 \w ( 1 + \w )  x_4^9 \big)  + 24 A ( 1 + \w )  x_4^6 
         ( -1 - 3 x_4^2 + \w ( x_4 + x_4^3 )  )
	  \Big),    \\
f^W_3 &=& \frac{( 1 + \w )}{24 x_4^4 {( -1 + x_4^2 ) }^4}  \Big( \tB \big( -144 + 108 ( 1 + 2 \w )  x_4 - 
         24 ( 5 + \w )  ( -3 + 2 \w )  x_4^2 \nonumber \\ && +
	 36 ( -11 + ( -16 + \w )  \w )  x_4^3 +  
         24 \w ( 29 + 7 \w )  x_4^4 - 36 ( -2 + \w ( 2 + 7 \w )  )  x_4^5 \nonumber \\ && + 
         3 ( -1 + \w )  ( -3 + 32 \w )  x_4^6 -
	 7 ( -1 + \w )  \w x_4^7  - 27 ( -1 + \w )
	  x_4^8 \nonumber \\ &&+  
         9 ( -1 + \w )  \w x_4^9 \big)  - 24 A ( -1 + \w )  x_4^6 
       ( -1 - 3 x_4^2 + \w ( x_4 + x_4^3 )  )
       \Big),   \\
f^W_4 &=& \frac{3 \tB {( -1 + \w x_4 ) }^2 ( 4 -
  10 x_4^2 + \w x_4 ( -1 + 7 x_4^2 )  ) }{x_4^4 {(
    -1 + x_4^2 ) }^4} .
\eea

Results including the $\tk^2$ corrections can be found at \cite{Website}.

\pagebreak

\section{Bulk geodesics}

To calculate the affine distance between the branes along a spacelike
geodesic we must solve the geodesic equations in the bulk.
Let us first consider the situation in Birkhoff-frame coordinates for which the
bulk metric is static and the branes are moving.  The Birkhoff-frame
metric takes the form \cite{TTS}
\[
\d s^2 = \d Y^2 - N^2(Y)\, \d T^2 + A^2(Y)\, \d \vec{x}^2,
\]
where for Schwarzschild-AdS with a horizon at $Y=0$,
\[
A^2(Y) = \frac{\cosh(2 Y/L)}{\cosh(2Y_0/L)}, \qquad N^2(Y) =
  \frac{\cosh(2 Y_0/L)}{\cosh(2Y/L)}\left(\frac{\sinh{(2Y/L)}}{\sinh{(2Y_0/L)}}\right)^2.
\]
At $T=0$, the $Y$-coordinate of the branes is represented by the parameter $Y_0$;
their subsequent trajectories $Y_\pm(T)$ can then be determined by integrating 
the Israel matching conditions, which read $\tanh{(2Y_\pm/L)}= \pm
\sqrt{1-V_\pm^2}\,$, 
where $V_\pm = (\d Y_\pm/\d T)/N(Y_\pm)$
are the proper speeds of the positive- and negative-tension branes respectively.
From this, it further follows that $Y_0$
is related to the rapidity $y_0$ of the collision
by $\tanh y_0 =\sech(2Y_0/L)$.

For the purpose of measuring the distance between the branes, a
natural choice is to use spacelike geodesics that are orthogonal to the four
translational Killing vectors of the static bulk, corresponding to
shifts in $\vec{x}$ and $T$. 
Taking the $\vec{x}$ and $T$ coordinates to be fixed along the
geodesic then, 
we find that $Y_{,\lambda}$ is constant for an affine parameter $\lambda$ along the geodesic.

To make the connection to our original brane-static coordinate system,
recall that the metric function $b^2(t,y) = A^2(Y)$, and thus
\[
Y_{,\lambda}^2 = \frac{(bb_{,t}t_{,\lambda}+b b_{,y} y_{,\lambda})^2}{b^4 -
  \theta^2} = n^2 (-t_{,\lambda}^2+t^2 y_{,\lambda}^2),
\]
where we have introduced the constant $\theta=\tanh{y_0}=V/c$.
Adopting $y$ now as the affine parameter, 
we have
\[
0 = (b_{,t}^2 b^2+n^2(b^4-\theta^2))t_{,y}^2 + 2 b_{,t}b_{,y}b^2 t_{,y}+(b_{,y}^2
b^2-n^2t^2(b^4-\theta^2)),
\]
where $t$ is to be regarded now as a function of $y$.

We can solve this equation order by order in $y_0$ using the series ansatz
\[
t(y) = \sum_{n=0}^\inf c_n y^n,
\]
where the constants $c_n$ are themselves
series in $y_0$.
Using the series solution for the background geometry given in Appendix
B, and imposing the boundary condition that $t(y_0)=t_0$, we obtain
\bea
c_0 &=&
t_0 + \frac{t_0\,y_0^2}{2} - 2\,t_0^2\,y_0^3 + \frac{\left( t_0 + 36\,t_0^3 \right) y_0^4}{24} - 
  t_0^2\left( 1 + 5\,t_0^2 \right) y_0^5 + \left( \frac{t_0}{720}
  + \frac{17\,t_0^3}{4} + 4\,t_0^5 \right) y_0^6 \nonumber \\
&& - \frac{t_0^2\left( 13 + 250\,t_0^2 + 795\,t_0^4 \right) y_0^7}{60} + O(y_0^8) \\
c_1 &=& 
2\,t_0^2\,y_0^2 + \left( \frac{5\,t_0^2}{3} + 5\,t_0^4 \right) \,y_0^4 - 8\,t_0^3\,y_0^5 + 
  \left( \frac{91\,t_0^2}{180} + \frac{23\,t_0^4}{6} + \frac{53\,t_0^6}{4}
  \right) \,y_0^6 + O(y_0^7) \\
c_2 &=& 
-\frac{t_0}{2} - \frac{t_0\left( 1 + 6\,t_0^2 \right) y_0^2}{4} + t_0^2\,y_0^3 - 
  \left( \frac{t_0}{48} - 2t_0^3 + 4t_0^5 \right)y_0^4 + \frac{\left( t_0^2 + 23\,t_0^4 \right) y_0^5}{2} + 
  O(y_0^6) \\
c_3 &=& 
-\frac{5\,t_0^2\,y_0^2}{3} - \frac{t_0^2\,\left( 25 + 201\,t_0^2 \right)
  y_0^4}{18} + O(y_0^5) \\
c_4 &=&
\frac{5\,t_0}{24} + \left( \frac{5\,t_0}{48} + \frac{7\,t_0^3}{4} \right) y_0^2 - \frac{5\,t_0^2\,y_0^3}{12} + 
  O(y_0^4) \\
c_5 &=&
\frac{61\,t_0^2\,y_0^2}{60} + O(y_0^3) \\
c_6 &=& 
-\frac{61\,t_0}{720} + O(y_0^2) \\
c_7 &=& 0 + O(y_0) .
\eea
Substituting $t_0=x_0/y_0$ and $y=\w y_0$ we find $x(\w)=x_0/y_0+O(y_0)$,
\ie to lowest order in $y_0$, 
the geodesics are trajectories of constant time lying solely along the
$\w$ direction.
Hence in this limit, the affine and metric separation of the branes
(defined in (\ref{d_m})) must necessarily agree.  
To check this, the affine distance between the branes is given by
\bea
\frac{d_a}{L} &=& \int_{-y_0}^{y_0}n\sqrt{t^2-t'^2}\,\d y \\
&=& 2\,t_0\,y_0 + \frac{\left( t_0 + 5\,t_0^3 \right) \,y_0^3}{3} - 4\,t_0^2\,y_0^4 + 
  \frac{\left( t_0 - 10\,t_0^3 + 159\,t_0^5 \right) \,y_0^5}{60} -
  \frac{2\,\left( t_0^2 + 30\,t_0^4 \right) \,y_0^6}{3} \nonumber \\
&& +  
  \frac{\left( t_0 + 31115\,t_0^3 - 5523\,t_0^5 + 12795\,t_0^7 \right) \,y_0^7}{2520} + O(y_0^8),
\eea
which to lowest order in $y_0$ reduces to
\[
\frac{d_a}{L} = 2\,x_0 + \frac{5\,x_0^3}{3} + \frac{53\,x_0^5}{20} +
\frac{853\,x_0^7}{168} + O(x_0^8) + O(y_0^2),
\]
in agreement with the series expansion of (\ref{d_m}).
(Note however that the two distance measures differ nontrivially at order
$y_0^2$).

To evaluate the perturbation $\delta d_a$ in the affine distance
between the branes, consider 
\bea
\delta \int \sqrt{\g \dot{x}^\mu \dot{x}^\nu} \d \lambda &=&
\frac{1}{2}\int \frac{\d \lambda}{\sqrt{g_{\rho\sigma} \dot{x}^\rho \dot{x}^\sigma}} \left(\delta\g
\dot{x}^\mu\dot{x}^\nu+g_{\mu\nu ,\kappa}\delta x^\kappa\dot{x}^\mu
\dot{x}^\nu +2 \g \dot{x}^\mu \delta \dot{x}^\nu\right) \nonumber \\
&=&\left[\frac{\dot{x}_\nu\delta x^\nu}{\sqrt{g_{\rho\sigma} \dot{x}^\rho
      \dot{x}^\sigma}}\right]+\frac{1}{2}\int\frac{\delta\g\dot{x}^\mu
  \dot{x}^\nu}{\sqrt{g_{\rho\sigma} \dot{x}^\rho \dot{x}^\sigma}}\,\d \lambda, 
\eea
where dots indicate differentiation with respect to the affine parameter
$\lambda$, and in going to the second line we have integrated by parts
and made use of the background geodesic equation
$\ddot{x}_\sigma=\frac{1}{2} g_{\mu\nu ,\sigma}\dot{x}^\mu \dot{x}^\nu$ and
the constraint $\g \dot{x}^\mu \dot{x}^\nu=1$. 
If the endpoints of the geodesics on the branes are unperturbed, this
expression is further simplified by the vanishing of the surface term.
Converting to coordinates where $t_0= x_0/y_0$
and $y= \w y_0$, to lowest order in $y_0$ the unperturbed geodesics
lie purely in the $\w$ direction, and so the perturbed affine distance
is once again identical to the perturbed metric distance (\ref{deltad_m}).

Explicitly, we find
\bea
&& \frac{\delta d_a}{L} =-\frac{2\,\left( B + A\,t_0^2 \right) \,y_0}{t_0} -
\left( \frac{B\,\left( 4 + 3\,t_0^2 \right)}{12\,t_0} +  
     \frac{A\,\left( t_0+ 9\,t_0^3 \right) }{3} \right) y_0^3 +
     \left( -4\,B + 4\,A\,t_0^2 \right) y_0^4\nonumber \\
&& - \left(\frac{B\,\left( 2 + 2169\,t_0^2 + 135\,t_0^4 \right)  + 
       2\,A\,t_0^2\,\left( 1 + 1110\,t_0^2 + 375\,t_0^4 \right)}
       {120\,t_0}\right)y_0^5 \nonumber \\
&& +  \left(\frac{4\,A\,t_0^2\,\left( 1 + 42\,t_0^2 \right)  -
       B\,\left( 4 + 57\,t_0^2 \right)}{6}\right) y_0^6 \nonumber
     \\
&& -   \Big(\frac{B\left( 4 + 88885t_0^2 + 952866t_0^4 + 28875t_0^6 \right)  + 
       4At_0^2\left( 1 - 152481 t_0^2 + 293517 t_0^4 +
       36015 t_0^6 \right)}{10080 t_0}\Big)\,y_0^7 \nonumber \\
&& +  O(y_0^8),
\eea
which, substituting $t_0=x_0/y_0$ and dropping terms of $O(y_0^2)$,  reduces to 
\[
\frac{\delta d_a}{L} = -\frac{2\,\tB}{x_0} - 2\,A\,x_0 - \frac{\tB}{4}\,x_0
  - 3\,A\,x_0^3 - \frac{9}{8}\,\tB\,x_0^3 - \frac{25}{4}\,A\,x_0^5 -  
  \frac{275}{96}\,\tB\,x_0^5 - \frac{343}{24}\,A\,x_0^7 + O(x_0^8),
\]
where $\tB=B y_0^2$.
Once again, this expression is in accordance with the series expansion
of (\ref{deltad_m}).  However, the perturbed affine and metric
distances do not agree at $O(y_0^2)$.

\end{appendix}

\bibliographystyle{apsrev}
\bibliography{finaldraft3}

\begin{thebibliography}{45}
\bibitem[{com({\natexlab{a}})}]{comment1}
\bibinfo{note}{Note that the radius of convergence of the series
  (\ref{W_series}) for $W(x)$ is $1/e$, and thus it converges for arguments in
  the range $-1/e\le x \le 0$ as required.}
\bibitem[{com({\natexlab{c}})}]{comment3}
\bibinfo{note}{In a companion paper \cite{cyclicpaper} we show that $A_4^L$ is
  in fact proportional to $\zeta$, the comoving curvature perturbation on the
  positive-tension brane, which is conserved on long wavelengths.}
\end{thebibliography}

\end{document}